\documentclass[twocolumn]{aastex63}

\usepackage{csquotes}
\usepackage{xcolor}

\newcommand{\HI}{H\,{\scriptsize I}}
\newcommand{\HII}{H\,{\scriptsize II}}
\newcommand{\CII}{[C\,{\scriptsize II}]}

\received{}
\revised{}
\accepted{}
\submitjournal{ApJ}

\shorttitle{H I in IC 63}
\shortauthors{. }
\graphicspath{{./}{figures/}}

\begin{document}

\title{High Resolution Observations of \HI\ in the IC 63 Reflection Nebula}

\correspondingauthor{L. Bonne, B-G Andersson}
\email{lbonne@usra.edu, bg@sofia.usra.edu}

\author[ 0000-0002-0915-4853]{Lars Bonne}
\affiliation{SOFIA Science Center/USRA, NASA Ames Research Center, M.S. N232-12, Moffett Field, CA 94035, USA}

\author[0000-0002-0786-7307]{B-G Andersson}
\affiliation{SOFIA Science Center/USRA, NASA Ames Research Center, M.S. N232-12, Moffett Field, CA 94035, USA}

\author[0000-0002-1261-6641]{Robert Minchin}
\affiliation{SOFIA Science Center/USRA, NASA Ames Research Center, M.S. N232-12, Moffett Field, CA 94035, USA}
\affiliation{National Radio Astronomy Observatory, P.O. Box O, Socorro, NM 87801, USA}

\author[0000-0002-6386-2906]{Archana Soam}
\affiliation{Indian Institute of Astrophysics, II Block, Koramangala, Bengaluru 560034, India}
\affiliation{SOFIA Science Center/USRA, NASA Ames Research Center, M.S. N232-12, Moffett Field, CA 94035, USA}

\author{Joshua Yaldaei}
\affiliation{Department of Physics, Santa Clara University, Santa Clara, 500 El Camino Real, Santa Clara, CA, USA} 

\author[0000-0002-1868-4485]{Kristin Kulas}
\affiliation{Department of Physics, Santa Clara University, Santa Clara, 500 El Camino Real, Santa Clara, CA, USA} 

\author[0000-0001-5996-3600]{Janik Karoly}
\affiliation{Jeremiah Horrocks Institute, University of Central Lancashire, Preston PR1 2HE, UK}
\affiliation{SOFIA Science Center/USRA, NASA Ames Research Center, M.S. N232-12, Moffett Field, CA 94035, USA}

\author[0000-0002-9342-9003]{Lewis B.G. Knee}
\affil{Herzberg Astronomy and Astrophysics Research Centre, National Research Council of Canada, 5071 West Saanich Road, Victoria, BC, V9E 2E7, Canada}

\author[0000-0002-5816-6623]{Siddharth Kumar}
\affiliation{Department of Physics, Indian Institute of Science, Bangalore 560012,India}

\author{Nirupam Roy}
\affiliation{Department of Physics, Indian Institute of Science, Bangalore 560012,India}



\begin{abstract}

Photodissociation regions (PDRs), where the (far-)ultraviolet light from hot young stars interact with the gas in surrounding molecular clouds, provide laboratories for understanding the nature and role of feedback by star formation on the interstellar medium.  While the general nature of PDRs is well understood - at least under simplified conditions - the detailed dynamics and chemistry of these regions, including gas clumping, evolution over time etc. can be very complex. We present interferometric observations of the 21 cm atomic hydrogen line, combined with \CII\ 158 $\mu$m observations, towards the nearby reflection nebula IC\,63. We find a clumpy \HI\ structure in the PDR, and a ring morphology for the \HI\ emission at the tip of IC 63. We further unveil kinematic substructure, of the order of 1~km~s$^{-1}$, in the PDR layers and several legs that will disperse IC 63 in $<$0.5 Myr. We find that the dynamics in the PDR explain the observed clumpy \HI\ distribution and lack of a well-defined \HI/H$_{2}$ transition front. However, it is currently not possible to conclude whether \HI\ self-absorption (HISA) and non-equilibrium chemistry also contribute to this clumpy morphology and missing \HI/H$_{2}$ transition front.


\end{abstract}

\keywords{(ISM:) photon-dominated region (PDR), techniques: interferometric}


\section{Introduction} \label{sec:intro}

The formation of high mass (OB) stars leads to stellar feedback in the form of radiation, stellar winds and eventually supernovae. These feedback processes ionize the surrounding interstellar medium (ISM) and can have an important impact on its dynamical and chemical evolution. At the interface of the fully ionized phase with the molecular phase this creates photodissociation regions (PDR) where FUV photons with energies between 6 eV and 13.6 eV dominate the chemistry \citep{Hollenbach1999,Roellig2007,Wolfire2022}. This leads to a chemical evolution as a function of depth into the cloud due to the decreasing amount of FUV photons. This chemical evolution includes, among (many) others, the transition from atomic to molecular hydrogen (H$_{2}$) and from ionized carbon (C$^{+}$) to carbon monoxide (CO) \citep[e.g.][]{Sternberg1995}. These different tracers can then also provide a view on the dynamics of the cloud as a function of depth in the PDR which helps to constrain the effect of stellar feedback on molecular cloud evolution \citep[e.g.][]{Schneider2020}. In addition, probing a variety of tracers allows to investigate the physical, dynamic and chemical structure of the PDR. Often, PDR models are based on plane parallel geometries \citep[e.g.][]{Tielens1985,Kaufman1999,LePetit2006}, yet PDR structure can be more complicated in the 3D turbulent ISM. Therefore, dedicated clumpy PDR models have been developed \citep[e.g.][]{Gierens1992,Stoerzer1996} as well as PDR codes, such as 3D-PDR \citep{Bisbas2012}, that make predictions by post-processing turbulent ISM simulations. In addition, it is possible that the evolution in some PDRs is driven by non-equilibrium chemistry due to a rapid progression of the ionization front which can particularly affect the atomic to molecular hydrogen (\HI/H$_{2}$) transition front \citep{Bertoldi1996,Stoerzer1998,Maillard2021}.\\\\
Here, we will investigate whether the atomic to molecular hydrogen transition in a PDR is governed by non-equilibrium chemistry or affected by the 3D turbulent dynamics of the region. For this we will particularly make use of observations of the \HI\ 21 cm line. However, because of the relatively long wavelength of this line, it is challenging to resolve the structure and dynamics of atomic hydrogen in PDRs with single dish observations even for the most nearby regions. Interferometric observations with incomplete u-v spacing coverage, on the other hand, risk resolving out the diffuse emission. To reliably probe the PDR of the reflection nebula/molecular cloud IC 63 in \HI\ we have combined observations of three radio interferometers and single dish data to ensure both high spatial and spectral resolution and full u-v spacing coverage.  We have combined observations from the Giant Metrewave Radio Telescope (GMRT), the Westerbork Synthesis Radio Telescope (WSRT), the Dominion Radio Astrophysical Observatory (DRAO) Synthesis Telescope (ST) \citep{Landecker2000} and the DRAO John A. Galt 26m single-dish telescope to probe the kinematics and structure of the PDR. Together these observations cover baselines from 23 km (with the GMRT) down to zero-spacing single dish data (DRAO) and provide the possibility to reach a resolution of 2.5\arcsec. 
In this study we present these interferometric \HI\ 21 cm observations of the IC 63 reflection nebula, complemented with the SOFIA \CII\ 158 $\mu$m observations from Caputo et al. (2023, submitted), as well as the H$_2$ data from \citet{Andersson2013} and \citet{Soam2021c}. Since carbon is still ionized when molecular hydrogen has already formed \citep{Tielens1985,Sternberg1995}, the combined data set allows a study of the PDR structure and dynamics as a function of depth in the PDR.\\\
IC\,63 is a nebula in the Sh2-185 \HII\ region which is illuminated and ionized by a B0 IV star, $\gamma$ Cassiopeia \citep{Karr2005}. This cloud is classified as type-B bright-rimmed cloud by \citet{Sugitani1991}. Gaia parallaxes provide a distance of $\lesssim 200$ pc to this region which makes it one of the closest \HII\ regions to the Sun \citep{soam2021b}. Because of its proximity, the interferometric observations reach a high spatial resolution ($\sim$1.5$\times$10$^{-2}$ pc). The nebula has also been extensively studied in many tracers, from the UV \citep[e.g.][]{witt1989,France2005} to the IR \citep[e.g.][]{Fleming2010,Andrews2018} and mm-waves \citep[e.g.][]{Jansen1994,soam2021b}.

The atomic hydrogen emission from IC 63 has previously been studied by e.g. \citet{Blouin1997} using the DRAO ST outside Penticton, British Columbia.  However, even with the 604.3m maximum fully sampled baseline of the DRAO ST, the angular resolution at 1.42 GHz of \HI\ data only reaches 1$^{\prime}\times$ 1.14$^{\prime}$ for IC 63 \citep{Blouin1997}. For comparison, the SOFIA/upGREAT \CII maps obtained by Caputo et al. (2023, submitted) have a beam size of $\sim$14\arcsec\ and the GBT/ARGUS CO(J=1-0) observations (in a forthcoming paper) have a beam size of 8\arcsec.  Therefore high sensitivity and high resolution observations of the region require more extensive arrays.

\section{Observations and Data Reduction} \label{sec:ObsRedux}
Because of the small angular size of the IC\ 63 PDR, very high spatial resolution is needed to analyze the nebula.  We therefore combine radio interferometry from the Giant Metrewave Radio Telescope (GMRT) in India, the Westerbork Synthesis Radio Telescope (WSRT) in the Netherlands and the Dominion Radio Astrophysical Observatory (DRAO) Synthesis Telescope (ST) in British Columbia, Canada. To ensure full u-v plane coverage, especially at short spacings, single dish observations using the DRAO 26m antenna were also added.

\subsection{GMRT Observations}

The GMRT H\,{\sc i} observations were taken in January 2017. These used the GMRT Software Backend \citep{Roy2010} with 512 channels across a bandwidth of 2.083~MHz, giving a channel separation of 0.86~km\,s$^{-1}$. The data were reduced and imaged using AIPS \citep{2003ASSL..285..109G} with a restoring beam of 8.5\arcsec. 

\subsection{WSRT Observations}

The WSRT H\,{\sc i} observations were taken over $4\times 12$ hours with the legacy WSRT system in the 36m+54m+72m+90m configurations in December 2008 using 1024 channels over a bandwith of 2.5~MHz, giving a channel separation of 0.52~km\,s$^{-1}$ and an angular resolution of 15.2\arcsec$\times$13.4\arcsec.
They were reduced and imaged using standard procedures in MIRIAD using the CLEAN algorithm \citep{1995ASPC...77..433S}. To check whether the clumpiness of the H\,{\sc i} in the resulting image was a CLEAN artifact, the data was also imaged using the MAXEN algorithm to carry out a maximum entropy deconvolution, resulting in a virtually identical image.

\subsection{DRAO Observations}
The DRAO \HI\ data utilized here were originally acquired as part of the Canadian Galactic Plane Survey \citep[CGPS][]{Taylor2003,Taylor2017}. Those observations were taken in Phase I of the CGPS in 1995--2000 and include single dish observations taken to fill in the zero-spacing data. The observations cover 272 channels with a channel separation of 0.82~km\,s$^{-1}$ and an angular resolution of 1$^{\prime}$x1.14$^{\prime}$.

\begin{figure}
    \centering
    \includegraphics[width=\hsize]{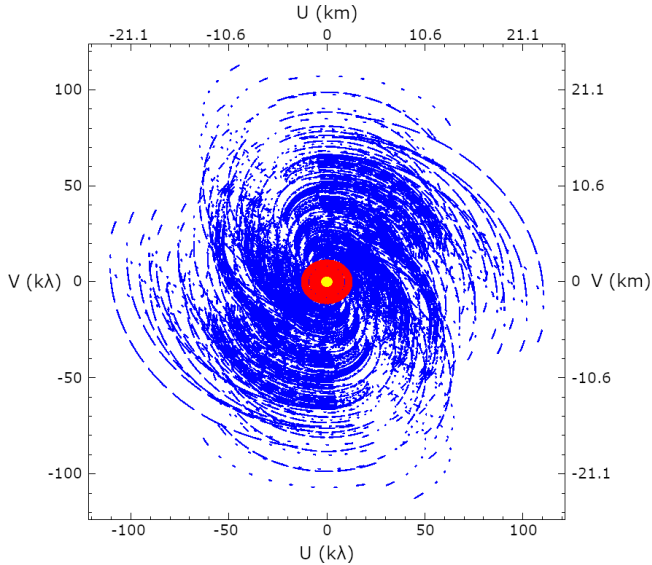}
    \caption{Coverage of the UV plane for the IC 63 observations by the different observatories. Blue is GMRT, red is WSRT and yellow is CGPS.}
    \label{fig:uvCoverage}
\end{figure}



\begin{deluxetable}{ccc}
\caption{Baselines for the different datasets}    \label{tab:baselines}
\tablehead{
 Dataset & Baselines
 }
 \startdata
 CGPS & 0--26\,m + 12.9--604.3\,m\\
 WSRT&36\,m--2.7\,km\\
 GMRT&100\,m--26\,km\\
 \enddata
\end{deluxetable}

\subsection{\HI\ data Reduction}

The three datasets were combined linearly in the Fourier domain using the {\sc immerge} task in MIRIAD \citep{1995ASPC...77..433S}, using the {\sc clean}ed and primary beam corrected WSRT and GMRT images and a UV range for relative flux calibration of 150 to 500m, which is present and well sampled in all three datasets (see Table \ref{tab:baselines}; Fig. \ref{fig:uvCoverage}). Prior to combination, the WSRT and DRAO images were regridded onto the spatial and spectral grid of the GMRT data. This gives a final combined dataset with the 8.5\arcsec angular resolution and the 0.86 km\,s$^{-1}$ spectral resolution of the GMRT data. In order to improve the S/N of the data and work with a resolution similar to the \CII\ data, we decided to proceed with a data cube smoothed to an angular resolution of 15\arcsec.

\subsection{SOFIA observations}

The \CII\ fine-structure line at 158 $\mu$m was observed with the upGREAT receiver \citep{Heyminck2012,Risacher2016,Risacher2018} onboard the Stratospheric Observatory for Infrared Astronomy (SOFIA) \citep{Young2012}. The data is part of SOFIA project 05\_0052 (PI: B-G Andersson). The IC 63 nebula was mapped in the total power on-the-fly mode (TP OTF), reaching an angular resolution of 15$^{\prime\prime}$. The observations were calibrated with the GREAT calibration software \citep{Guan2012} and the main beam efficiencies for the 7 different pixels of upGREAT vary between 0.59 to 0.68. To create the final data cubes, a second order baseline was fitted to the data with CLASS\footnote{https://www.iram.fr/IRAMFR/GILDAS/doc/html/class-html/class.html} and the data was spectrally smoothed to a resolution of 0.4 km s$^{-1}$. Caputo et al. (2023, submitted) will present a more detailed description of the \CII\ observations and data reduction.


\begin{figure*}
    \centering
    \includegraphics[width=0.4\hsize]{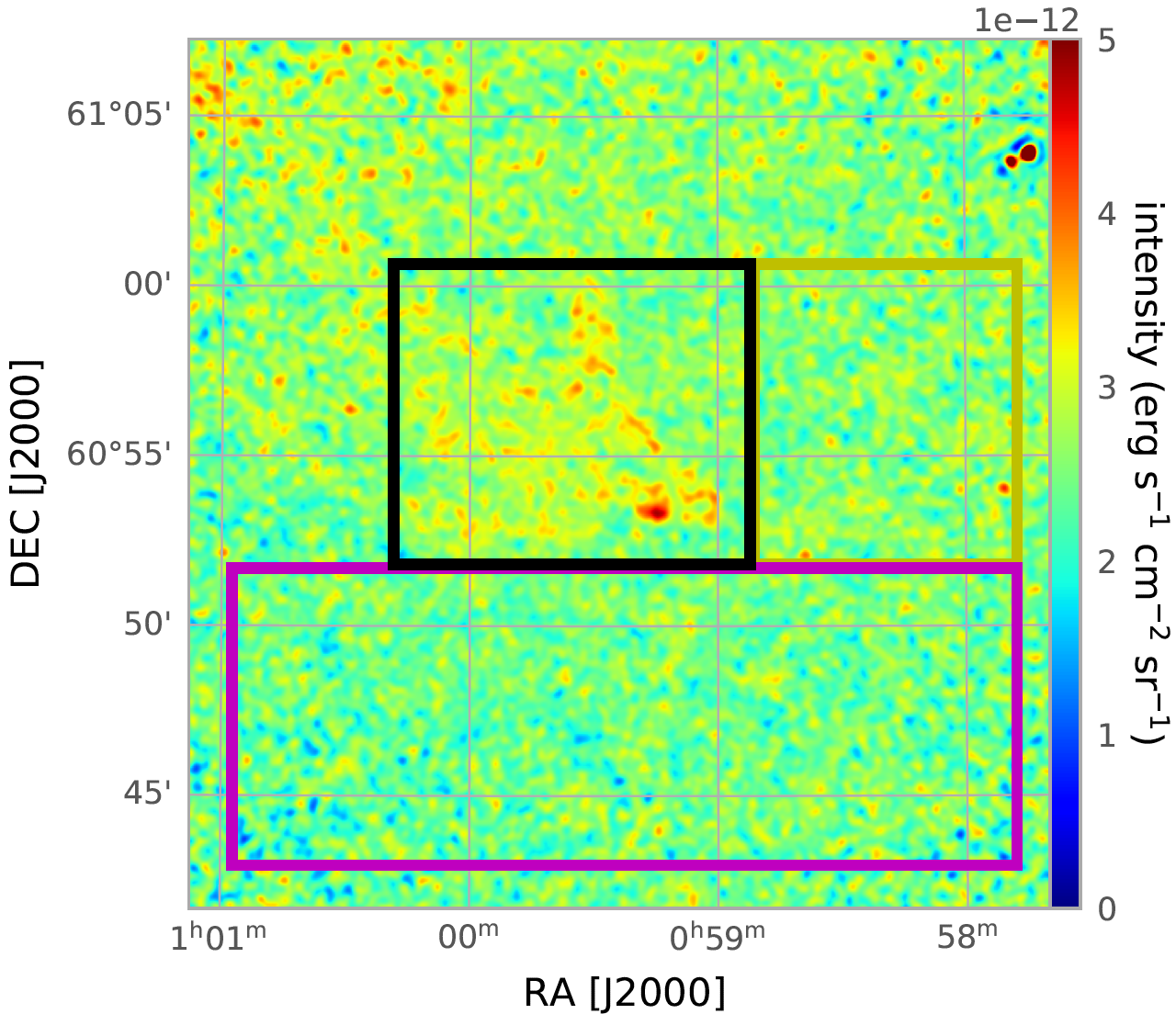}
    \includegraphics[width=0.51\hsize]{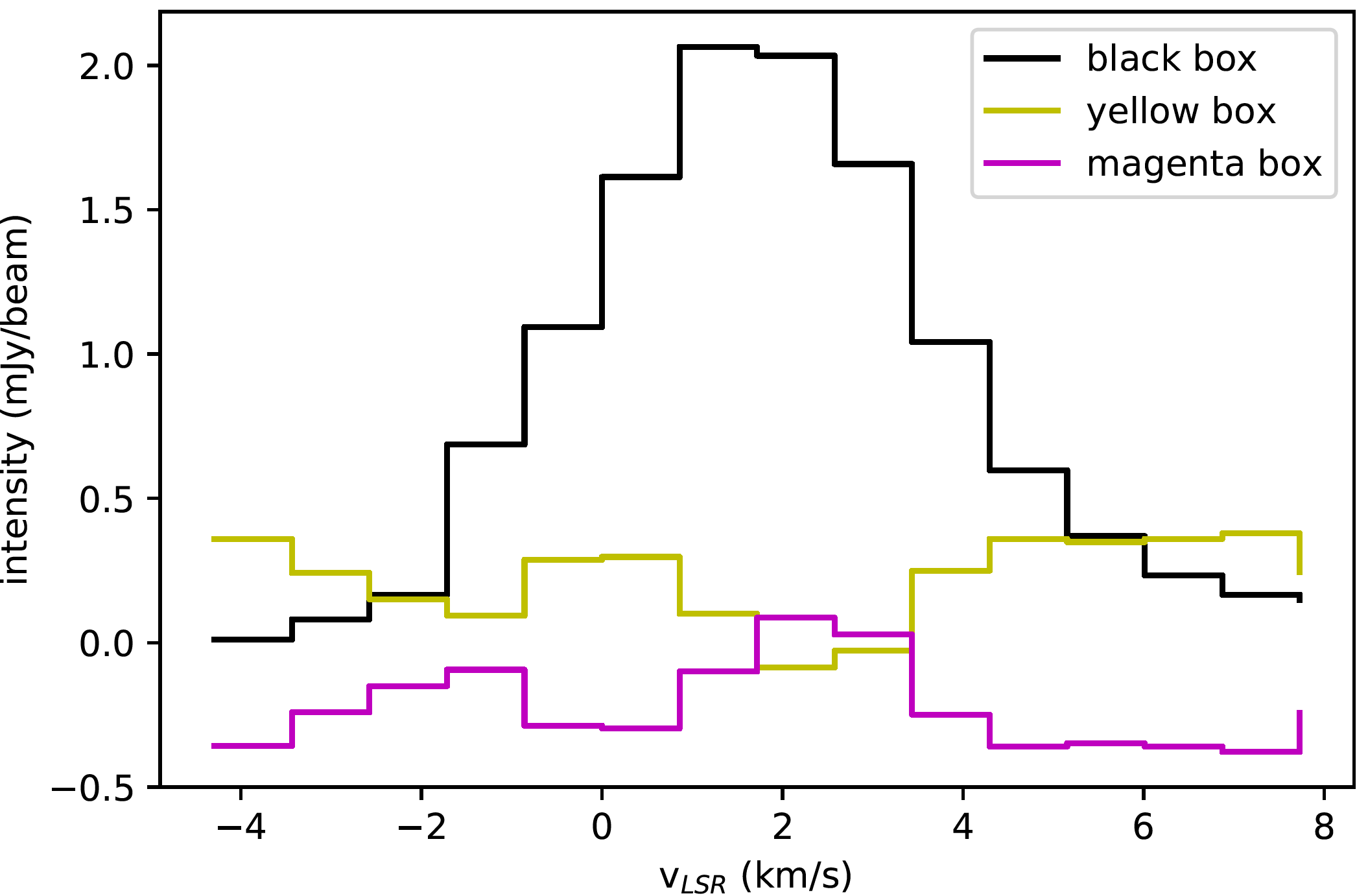}
    \caption{{\bf Left}: \HI\ integrated intensity map (without background subtraction) of the IC 63 region. The yellow and magenta boxes indicate the regions used to make an average spectrum for background subtraction. The black box indicates the region associated with IC 63. {\bf Right}: The spectra from the corresponding colored boxes on the left after background subtraction. The emission from IC 63 (black) has a clearly defined profile while the background regions (yellow \& magenta) are relatively close to the baseline.}
    \label{fig:backgroundSubtraction}
\end{figure*}

\section{Results} \label{res}

\subsection{Galactic background removal}
Since IC 63 is located relatively close to the Galactic plane (b $\approx$ -2$^{o}$), the observed \HI\ spectra are heavily confused/contaminated by background emission \citep{Taylor2003,McClureGriffiths2005}. In order to obtain the actual emission from IC 63, the galactic background has to be removed. To do so, we take a relatively simple approach where we define an average \HI\ background spectrum from regions of the data cube that do not contain emission associated with IC~63. We chose this approach because IC 63 is already clearly visible in the available data cube and the method gives excellent control on what is removed to isolate the emission associated with IC~63.\\
Two rectangular regions were used to define the average background spectrum for IC 63. These two regions are indicated by the yellow and magenta boxes presented in Fig. \ref{fig:backgroundSubtraction}. They are the only two regions that do not contain emission from IC 63 (which is located in the black box in Fig. \ref{fig:backgroundSubtraction}). For the rest of the map, i.e. the north-east of the map, it is not certain whether there is a contribution from IC~63 or IC~59 in the north-west corner of the map. As a result, these other regions are not taken into account for the background subtraction. Subtracting the average background spectrum in each spatial pixel of the full data cube then creates the data cube that contains the emission associated with IC~63. The resulting \HI\ integrated intensity map is shown in Fig. \ref{fig:mom0map} and the resulting average spectrum for IC 63 is presented in the right panel of Fig. \ref{fig:backgroundSubtraction}. This shows a Gaussian line profile. From this point forward, we will work with the background subtracted data unless mentioned otherwise.

\subsection{The integrated intensity map}
\begin{figure*}
    \centering
    \includegraphics[width=0.47\hsize]{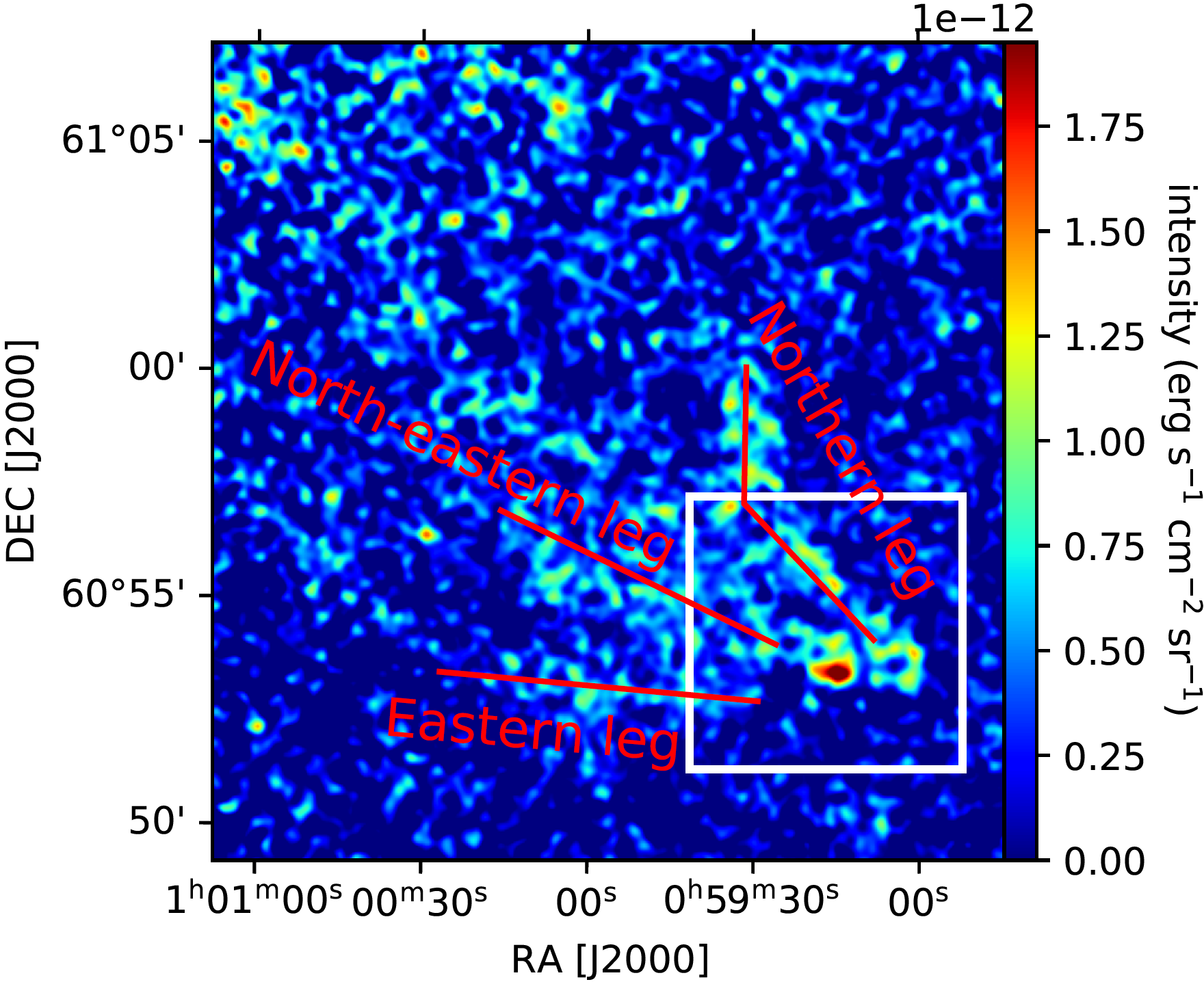}
    \includegraphics[width=0.45\hsize]{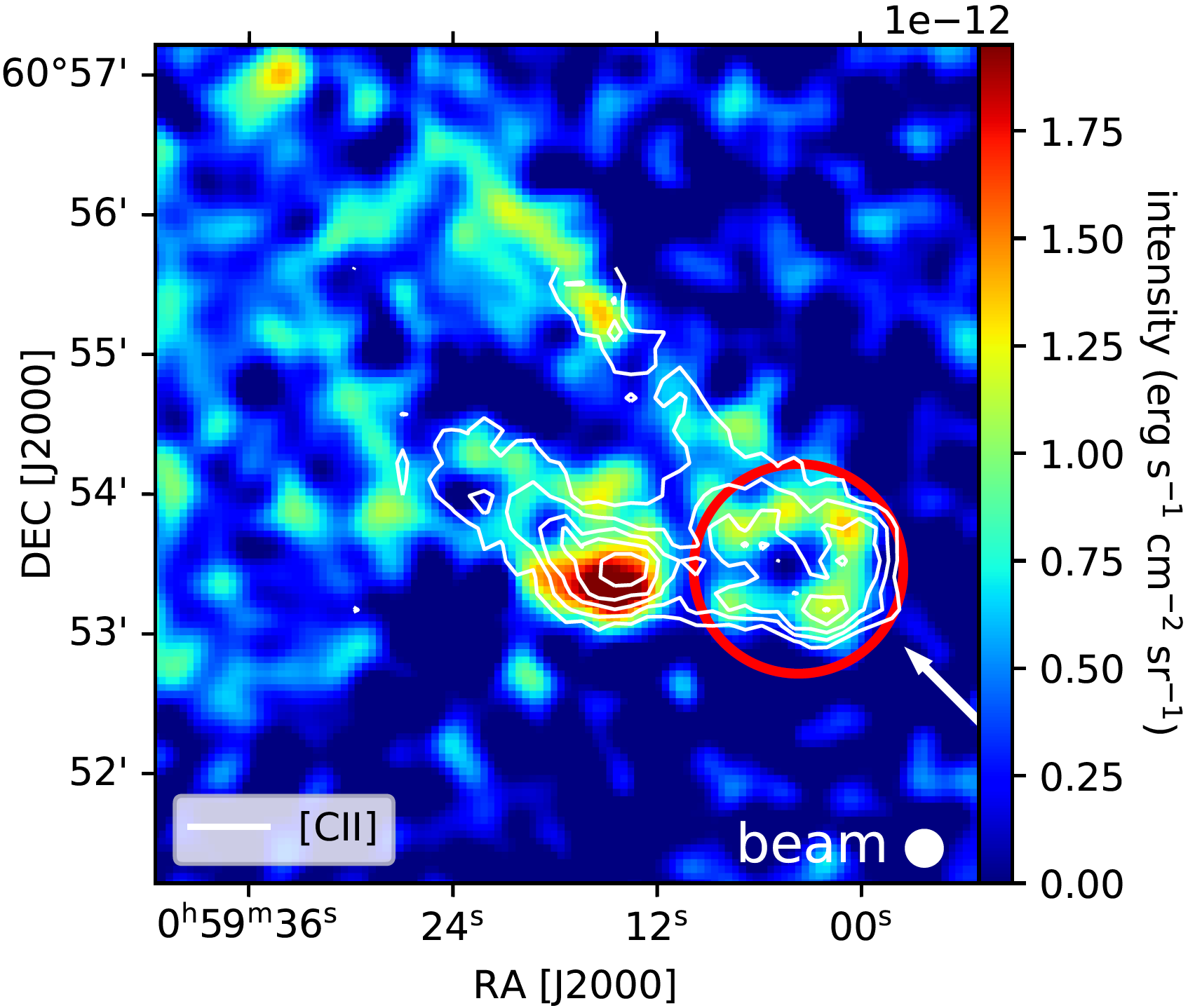}
    \caption{{\bf Left}: The \HI\ intensity map of IC 63 that is obtained from the background-subtracted data cube. The white box outlines the zoomed-in region presented on the right. The red lines indicate the three legs visible in \HI\ that are connected to the tip of IC~63 with their names. {\bf Right}: A zoom into the brightest region of \HI\ intensity map with the \CII\ intensity contours overlaid (starting at 10$^{-4}$ erg s$^{-1}$ cm$^{-2}$ sr$^{-1}$ with increments of 5$\times$10$^{-5}$ erg s$^{-1}$ cm$^{-2}$ sr$^{-1}$). The red circle highlights the circular morphology seen in \HI\ towards the tip of IC 63. The white arrow indicates the direction of the incident radiation from $\gamma$ Cas. The full white circle indicates the 15$^{\prime\prime}$ beam size of the \HI\ data.}
    \label{fig:mom0map}
\end{figure*}

The integrated intensity map of the \HI\ emission towards IC 63 is presented in Fig. \ref{fig:mom0map}. Overall, it shows the same morphology for IC 63 that is also seen in e.g. the WISE maps \citep[e.g.][]{soam2017}. When looking in more detail, it shows a bright clump and a ring-like morphology at the south-west head of the region with further \HI\ emission in elongated structures pointing towards the north, north-east and east of the map (which we here also call $legs$). The \HI\ emission in these legs show a quite pronounced clumpy morphology as well. The \CII\ map is more limited in size and only covers the head of IC~63. Inspecting the \CII\ emission in Fig. \ref{fig:mom0map} shows that the brightest emission is located towards the head of the nebula with lower brightness emission extending further along the start of the legs in the region. 
It is interesting to note that there are regions in Fig. \ref{fig:mom0map} with \CII\ emission that do not have a counterpart in \HI\ emission.\\ 
We do not expect that the clumpiness of the \HI\ emission is the result of filtering during the data reduction. Reducing the observations using the CLEAN technique, which tends to create clumpy structures, and the MEM technique, which produces smoother structures, we find that the clumpy structure of the \HI\ emission is maintained in both. This is shown in App. \ref{sec:CLEANvsMEM}. Inspecting the zoom into the tip of IC 63 in Fig. \ref{fig:mom0map}, we also find that the \HI\ emission shows a remarkable ring-like feature with an intensity drop in the middle. This behaviour is not seen in \CII, see Fig. \ref{fig:mom0map}, which rather shows bright emission towards the front of the PDR followed by a gradually decreasing intensity towards the back.\\ 
Given the relatively limited signal-to-noise ratio (S/N) of the \HI\ observations, it could be suggested that the clumpiness of the \HI\ data is simply due to the noise. To assess this, we determined the noise in the lower region of the integrated map without any known emission (i.e. the yellow and magenta boxes in Fig. \ref{fig:backgroundSubtraction}). This indicates a noise rms of 3.3$\times$10$^{-13}$ erg s$^{-1}$ cm$^{-2}$ sr$^{-1}$. However, inspecting the map we identify several clumps/locations, including the ring, that have a contrast above 10$^{-12}$ erg s$^{-1}$ cm$^{-2}$ sr$^{-1}$ (i.e. higher than 3$\sigma$) compared to their surroundings. Even though the limited S/N in the integrated map might have some effect, we find that several clumps should not be a S/N effect and therefore we are relatively confident that the \HI\ emission really is rather clumpy. More sensitive observations in the future should allow to explore this clumpiness with greater confidence.

\subsection{The kinematics of IC 63}
\begin{figure*}
	\centering
    \includegraphics[scale=0.4,angle=0]{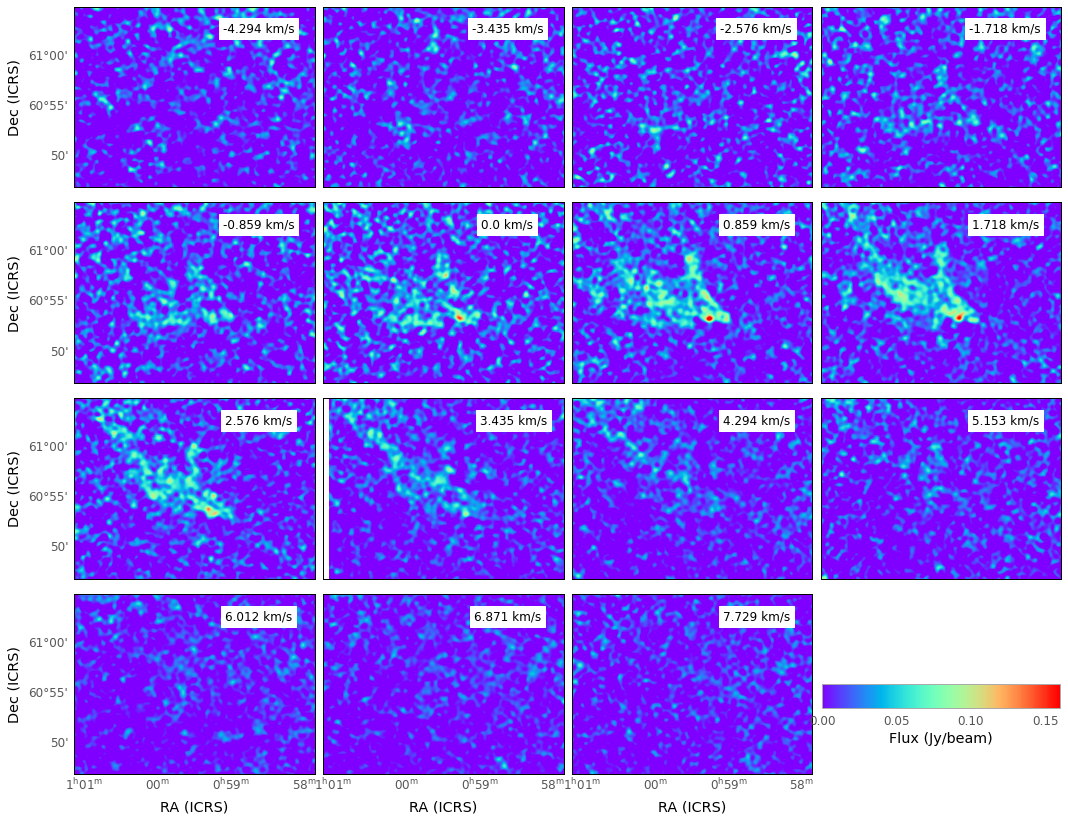}
	\caption{Channel maps between -4.3 and 7.7 km s$^{-1}$ of the IC 63 region in steps of $\sim$ 0.8 km s$^{-1}$.}
	\label{fig:chanMapHI}
\end{figure*}

\begin{figure*}
	\centering
	\raisebox{-0.5\height}{\includegraphics[width=0.37\hsize]{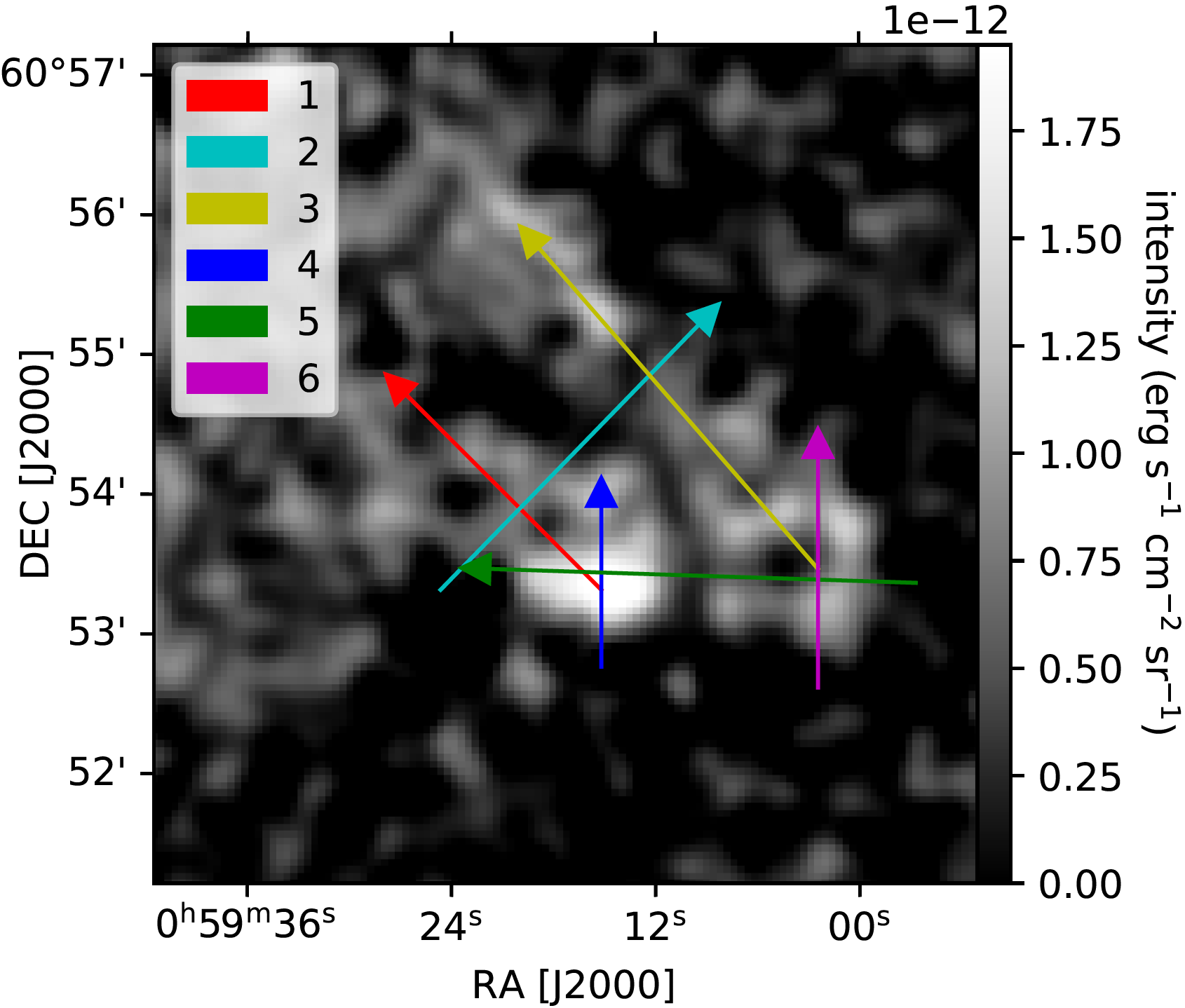}}
    \raisebox{-0.5\height}{\includegraphics[width=0.6\hsize]{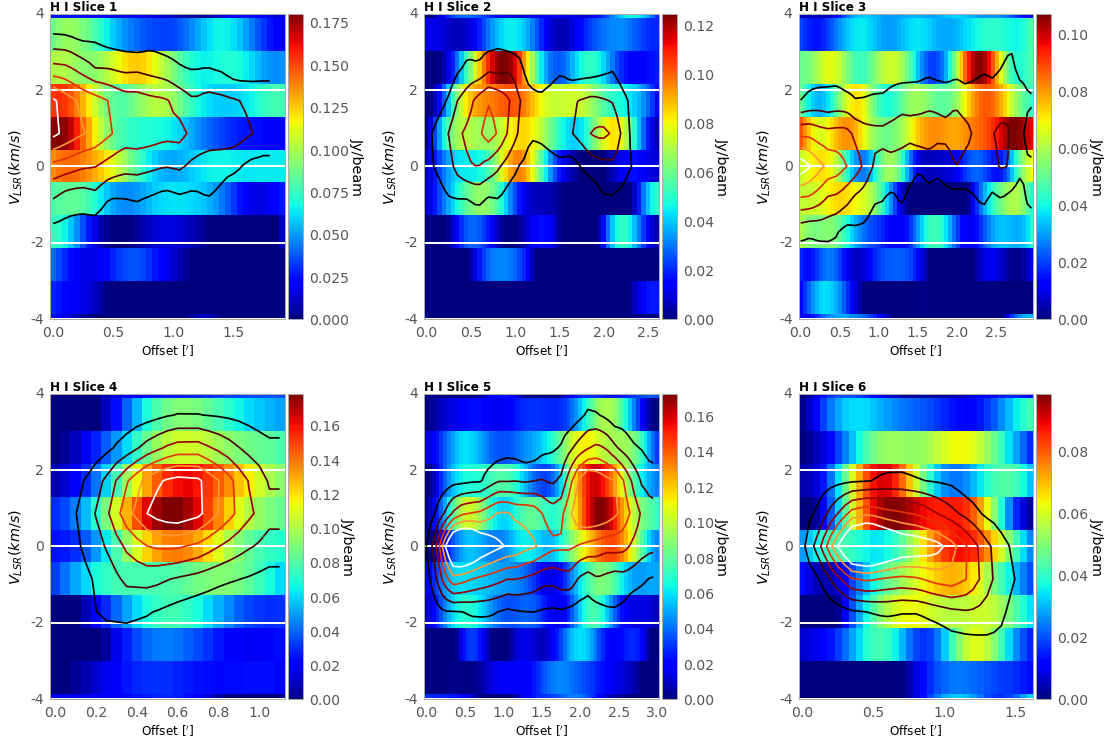}}
	\caption{{\bf Left}: The \HI\ moment zero map for IC 63 with the colored arrows indicating the cuts that were made to construct position-velocity (PV) diagrams on the right. {\bf Right}: The \HI\ PV diagrams along the axes/arrows indicated in the figure on the left. The contours indicate the corresponding \CII\ PV diagrams starting at 2 K with increments of 2 K. The darkest contours indicate the weakest emission and the white contour indicates the brightest emission.}
	\label{fig:pvDiagrams}
\end{figure*}

\begin{figure}
    \centering
    \includegraphics[width=\hsize]{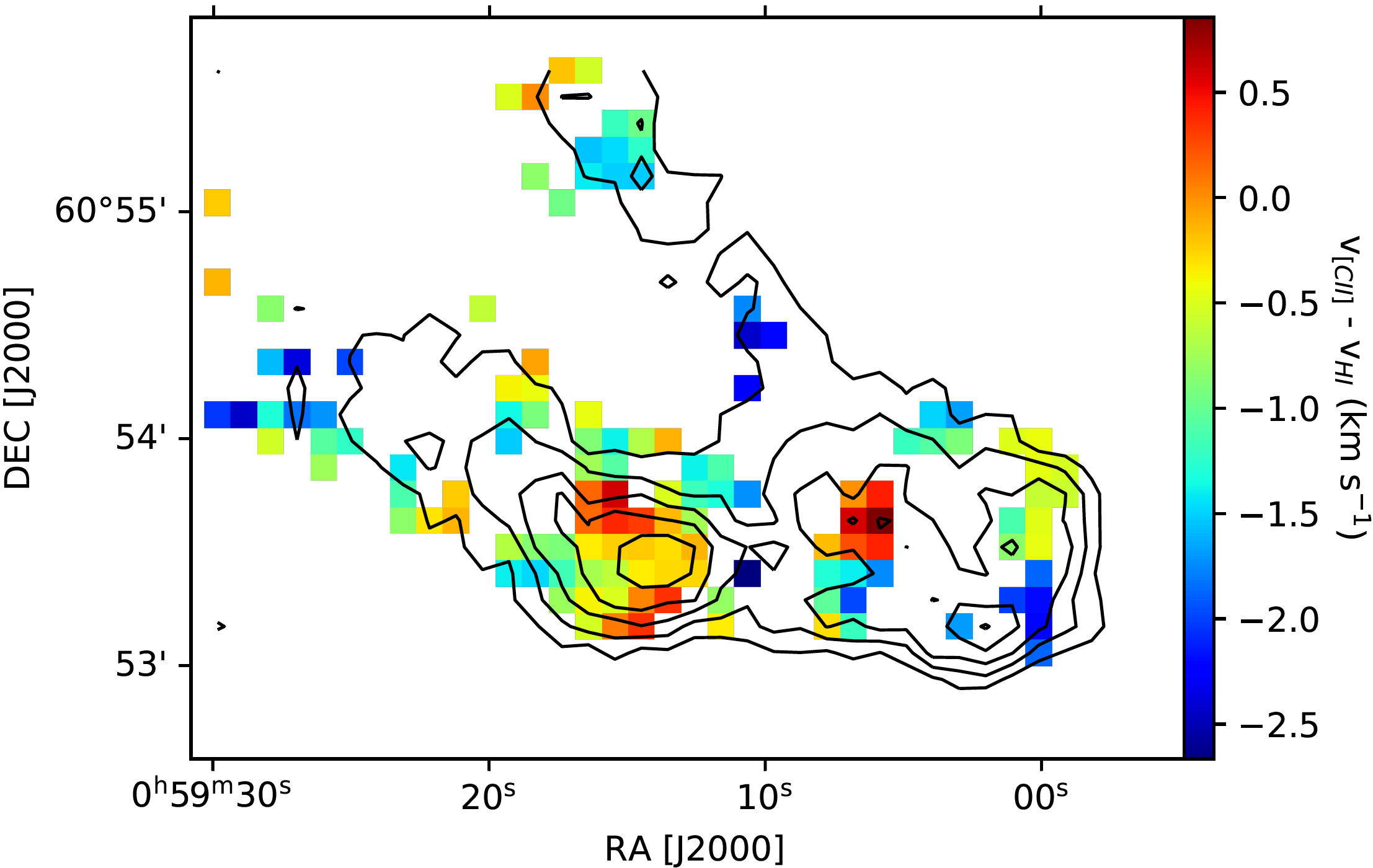}
    \caption{The difference in central velocity of the emission in \CII\ (Caputo et al. 2023, submitted) and \HI\ at the locations where both lines are detected. The black contours indicate the integrated \CII\ emission starting at 10$^{-4}$ erg s$^{-1}$ cm$^{-2}$ sr$^{-1}$ with increments of 5$\times$10$^{-5}$ erg s$^{-1}$ cm$^{-2}$ sr$^{-1}$.}
    \label{fig:velDiffMap}
\end{figure}

The channel maps in Fig. \ref{fig:chanMapHI} provide a view of the \HI\ kinematics in IC 63. At the most blueshifted velocities ($\lesssim$ 0 km s$^{-1}$), the most eastern part of IC 63 is  prominent. At slightly more redshifted velocities (0.8-1.7 km s$^{-1}$), the northern leg becomes bright with the north-eastern leg between 0.8 and and 3.4 km s$^{-1}$. This clearly demonstrates that the different legs of IC 63 are velocity-coherent gas structures in the region.\\
For a more detailed look into the kinematics of IC 63, we also made several position-velocity (PV) diagrams that cut through the IC 63 region. These PV diagrams go along and perpendicular to north-eastern and northern legs, and across the ring and brightest \HI\ clump. The resulting PV diagrams, overlaid with \CII\ contours are presented in Fig. \ref{fig:pvDiagrams}. Focusing first on the \HI\ emission alone, it is observed that there is a velocity gradient from more blueshifted to more redshifted along the northern leg (PV 3) and towards the north-east from the tip (PVs 5 + 1). In the cut perpendicular to the legs (PV 2), a hint of the velocity gradient west to east is seen in \HI. It is however tentative due to the limited spectral resolution and relatively high noise in the \HI\ data. Lastly, the cut through the ring at the tip of the nebula shows a velocity gradient from south (1-2 km s$^{-1}$) to the north (0 km s$^{-1}$) (see PV 6) while the clump to the east does not show any clear gradient (see PV 4).\\\\
When comparing the \HI\ and \CII\ kinematics -the \CII\ kinematics are presented and studied in more detail in Caputo et al. (2023, submitted)- both differences and similarities are observed. The kinematics of the legs in \HI\ and \CII\ appear to show the same behaviour, see Fig. \ref{fig:pvDiagrams}, even though some regions that are detected in \CII\ are not detected in \HI\ which is the result of the clumpy \HI\ structure. The velocity gradient from west to east that was tentatively observed in the \HI\ channel maps is not really clear in \CII\ (see PV 2 of Fig. \ref{fig:pvDiagrams}). However, this could be the case because this velocity gradient was tentative from the start. Most remarkable is probably the difference of the centroid velocity for \HI\ and \CII\ towards the head of IC 63. Fig. \ref{fig:pvDiagrams} shows that the \CII\ velocity field remains constant towards the ring whereas a gradient is observed in the \HI\ emission. This velocity difference of both lines is confirmed when fitting the \CII\ and \HI\ data with a single Gaussian velocity profile, which finds velocity difference of the order of 1-2 km s$^{-1}$, see Fig. \ref{fig:velDiffMap}. From this figure it is observed that \CII\ is predominantly blueshifted with respect to \HI\ in most parts of the ring. On the other hand, \CII\ is slightly redshifted with respect to \HI\ in the bright clump to the east of the ring (in particular at the edge of the clump) and in the eastern region of the ring itself. Note that there is some uncertainty on the fitted central velocities because of the low S/N of the \HI\ data. However, the velocity typically has an uncertainty of 0.1 km s$^{-1}$ and does not go over 0.2 km s$^{-1}$. The velocity shifts thus appear to be statistically significant. 

\section{Discussion} \label{disc}

\subsection{The dynamic evolution and timescales of IC 63}
The three major $legs$, defined in Fig. \ref{fig:mom0map}, of IC 63 are each seen at a different velocity, confirming they are coherently moving substructures of IC 63. From the \HI\ channel maps it was observed that the most eastern leg is the most blueshifted. This leg is proposed to be the closest to us \citep{Andersson2013}, which would imply that the different legs are currently being dispersed. Based on the 0.5 pc physical size of the \HI\ legs at a distance of $\sim$ 170 pc \citep{vanLeeuwen2007}, the velocity difference of $\sim$1-2 km s$^{-1}$ implies that IC 63 might be fully dissolved in $\le$ 0.5 Myr. Interestingly, this is similar, and even slightly shorter, than the dispersal timescales that are found with \CII\ for denser regions around high-mass star formation \citep[][]{Tiwari2021,Bonne2022}. As IC 63 is likely at a more evolved stage where most of the dense gas has already been dispersed, this indicates that the more diffuse gas around star forming regions can be dispersed on relatively similar timescales. Another interpretation for this velocity field could be a relatively rapid rotational motion of the legs if IC~63 is spinning, although it is unclear what would drive the spinning of this region. It is also observed that the legs to the north of the map are found to be generally more redshifted in Fig. \ref{fig:chanMapHI}. This velocity gradient could be associated with the dispersal of the region, but it is difficult to reach a firm conclusion.\\
Based on Fig. \ref{fig:velDiffMap} we noted that \HI\ and \CII\ show up at slightly different velocities, in particular towards the bright clump and ring at the head of IC 63. As \HI\ is expected to trace the outer regions of the PDR compared to \CII, because H$_{2}$ can form while carbon is still ionized \citep{Tielens1985,Sternberg1995}, this suggests that there is a velocity gradient over the PDR. The bright PDR in IC 63 is thus a dynamic structure with local velocity differences up to 1 km s$^{-1}$. As these clumps have a size of only 0.05 pc, the structure of these bright PDRs would thus be changing on a relatively short timescale of only $\sim$ 5$\times$10$^{4}$ yr. The potential effect of these dynamics on the chemical structure of the PDR will be discussed in more detail in the next section.

\subsection{The ring and clumpy PDR structure of \HI\ in IC 63}
\begin{figure*}
\begin{center}
\includegraphics[width=0.45\textwidth]{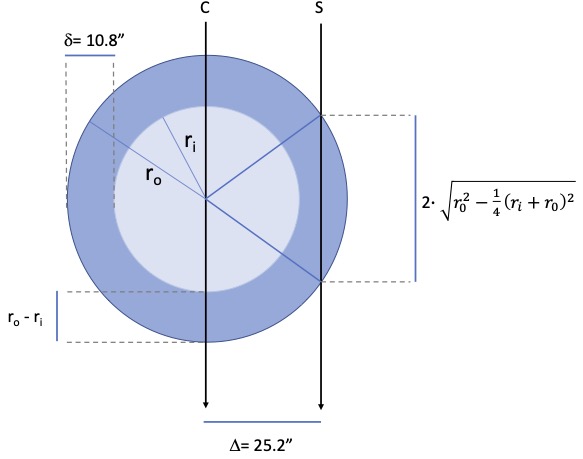}
\includegraphics[width=0.24\textwidth]{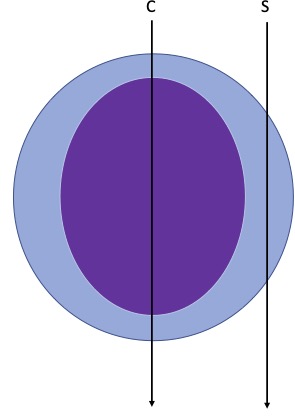}
    \caption{\textbf{Left:} The geometry of a simple model to estimate whether the observed \HI\ ring is the projection of a spherical ring (see Eqs. \ref{equ:fractPeak} \& \ref{equ:fract}). In this specific schematic presentation, S corresponds to the path length through the center of the shell (S$_{cent}$).  If we assume that all of the \HI\ emission is optically thin, we can compare the geometric ratio of ring size and rim/shell thickness with the ratio of the related brightness temperatures.  The observed ring radius is $\Delta$=25.2\arcsec\ and the shell thickness (FWHM) yields $\delta$=22.1\arcsec. \textbf{Right:} The observed intensity ratio could be recovered from this  model if we assume that the shell is anisotropic, and/or that the center of the region is filled with cold, self-absorbing (HISA), gas.}
    \label{Fig:ring_cartoon}
    \end{center}
\end{figure*}

\begin{figure}
	\centering
    \includegraphics[width=\hsize]{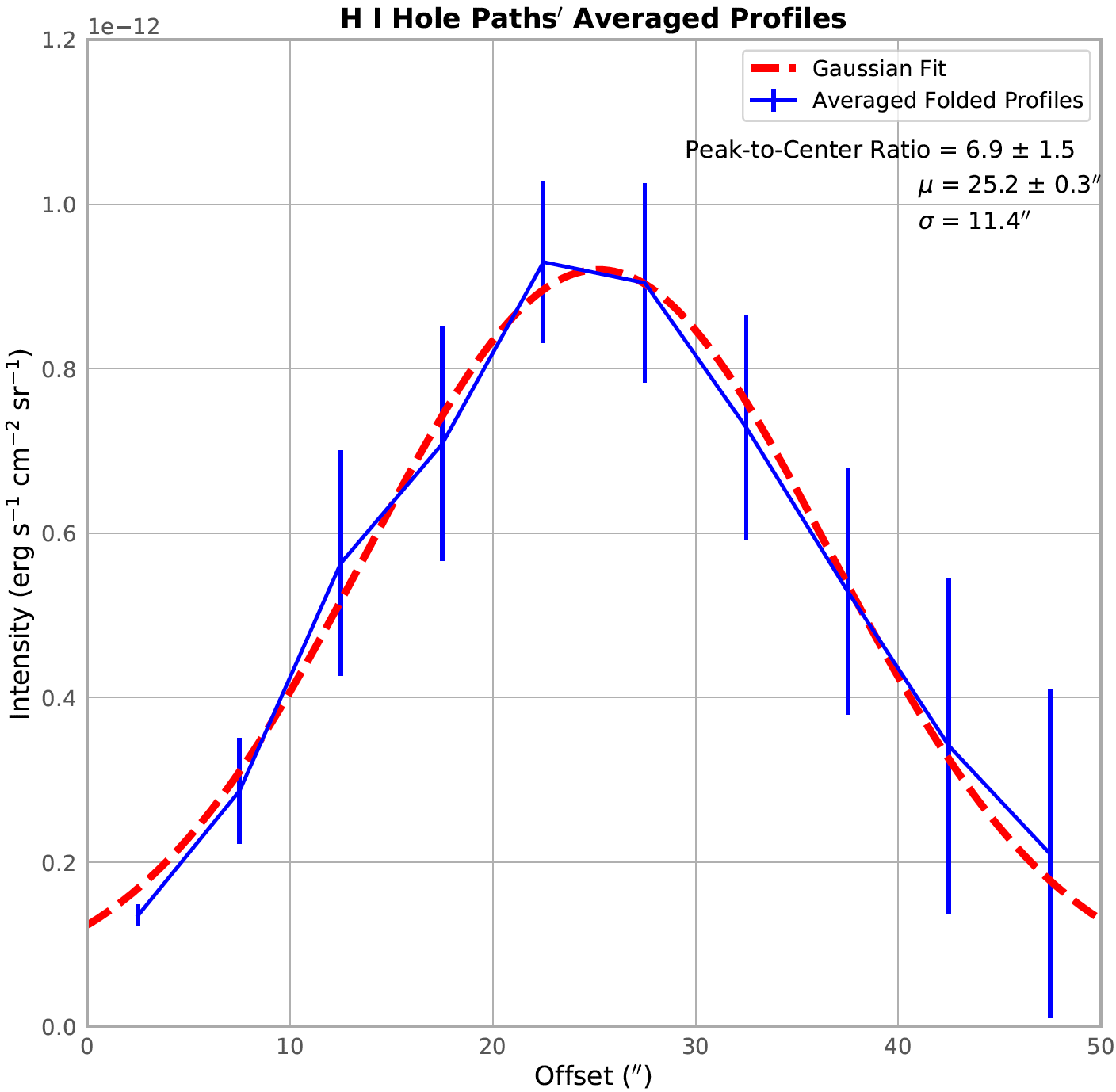}
	\caption{The average \HI\ emission profile of the ring as a function of the offset from the center of the ring with a Gaussian profile fitted to it (red dashed line). The error bars indicate the standard deviation of the emission over the ring at each offset.}
	\label{fig:avHIprofile}
\end{figure}

In order to assess the emission profile of the \HI\ ring-like structure at the tip of the nebula, we created 8 radial cuts through the ring. The resulting \HI\ profiles as a function of radius are presented in App. \ref{sec:indHIprofiles}. This confirms an axial symmetry of the emission with the peak emission in the 8 cuts only varying within 30\% among the different profiles, which is expected based on the noise rms. The dip towards the center of the ring is however larger than 3$\times$ the noise rms. In App. \ref{sec:indHIprofiles} it is also verified that the observed ring-like morphology is retrieved from the data of the different interferometers. {With this evidence of a ring-like structure}, we folded these 8 \HI\ profiles (i.e. average the profile at negative offset with the profile at positive offset) and then averaged all 8 profiles into a single \HI\ emission profile for the ring. This average emission profile, fitted with a Gaussian distribution, is presented in Fig. \ref{fig:avHIprofile}. The central intensity is 1.4$\times$10$^{-13}$ erg s$^{-1}$ cm$^{-2}$ sr$^{-1}$ and the peak intensity of the ring is 9.3$\times$10$^{-13}$ erg s$^{-1}$ cm$^{-2}$ sr$^{-1}$, resulting in a peak to center ratio of $\sim$7. The question arises whether this ring-like structure is the result of self-absorption or rather a result of the hydrogen chemistry or dynamics in the PDR. Investigating the individual spectra for self-absorption unfortunately does not provide conclusive insight because of the limited S/N for the currently available observations. We can however constrain this possibility, as well as chemical effects, with a toy model. This model assumes a spherical geometry for the optically thin atomic hydrogen that is organized in a shell with a constant spin temperature, see Fig. \ref{Fig:ring_cartoon}. In the interior of the shell we consider either the lack of \HI\ emission due to chemistry, e.g. the formation of H$_{2}$ which can be supported by the H$_{2}$ fluorescence observations in \citet{Andersson2013}, or the presence of cold atomic \HI\ self-absorbing (HISA) gas. With this toy model we can use the observed size and thickness of the ring, and compare it to the ratio of brightness temperature at the center and rim. We can also evaluate whether the required \HI\ absorbing layer is consistent with the A$_{V}$ (i.e. extinction in the V band) of IC 63.

If we assume that the shell is basically devoid of \HI, due to the formation of H$_{2}$, then the ratio at the center and the rim will be proportional to the path lengths through the shell, see Figure \ref{Fig:ring_cartoon}.  If we designate the inner and outer radii of the shell r$_i$ and r$_o$, the radial path length at the center of the shell is simply C=2$\cdot$(r$_o$ - r$_i$) and the distance from the center to the middle of the rim is $\frac{1}{2}\cdot(r_o + r_i)$. The path length through the middle of the shell is then S$_{cent}$=2$\cdot\sqrt{r_o^2-0.25\times(r_o + r_i)^2}$. For further use in equations we here define the size ($\Delta$) and thickness ($\delta$) with the following relations:

\begin{equation}
\delta=r_o - r_i; 
\Delta=\frac{1}{2}\cdot(r_o + r_i)
\end{equation}

The largest pencil-beam column density through the shell occurs for at r=r$_i$ with a resulting path length ratio
\begin{equation}\label{equ:fractPeak}
    \frac{S_{long}}{C}= \frac{\Delta + \frac{\delta}{2}}{\delta}\sqrt{1 - \left( \frac{\Delta - \frac{\delta}{2}}{\Delta + \frac{\delta}{2}} \right)^{2}}
\end{equation}
However, we are comparing to observations with a finite-sized beam, and the predicted shell intensity profile is not symmetric at the peak \citep[see Fig. 14 in][]{Kabanovic2022}. This can thus affect the observed S/C ratio and slightly shift the radius of the observed peak emission in a non-trivial way. Therefore, we also use the center path length through the shell which gives

\begin{equation}\label{equ:fract}
    \frac{S_{cent}}{C}=\frac{\sqrt{(\Delta+\frac{\delta}{2})^2 - \Delta^2}}{\delta}
\end{equation}
This provides an expected range of observed S/C ratios that accounts for the fact that we are observing at finite resolution.

Figure \ref{fig:avHIprofile} shows that $\Delta\approx$25.2\arcsec and $\sigma_{\delta}\approx$11.4\arcsec ($\sigma_{\delta}$ is the observed width of the shell). Deconvolving the shell-width with the 6.4\arcsec\ beam width (= FWHM$_{beam}$/2$\sqrt{2\text{ln}2}$ with FWHM$_{beam}$ = 15\arcsec), $\sigma_{\delta}$ becomes 9.4\arcsec.  The width of the shell is probably better approximated by the FHWM, rather than the Gaussian width $\sigma$ which yields $\delta\approx$22.1\arcsec.  Inserting these values into Eq. \ref{equ:fractPeak} and \ref{equ:fract}, we find path length ratios S/C in the range of 1.2 to 1.5. However, the observed brightness temperature ratio for the shell and center of the ring is T$_S$/T$_C$=6.9$\pm$1.5, which is significantly larger than the geometric ratio. It might thus be that the \HI\ shell is not spherical, see e.g. Fig. \ref{Fig:ring_cartoon}, or that there is HISA at the center of the ring.\\\\

To calculate the potential cold \HI\ column density, and compare it to the visual extinction in the cloud, we calculate the HISA optical depth
\begin{equation}
    \tau_{HISA}(v) = -{\rm ln}\left(1 - \frac{T_{on-off}(v)}{T_{HISA} - pT_{off}(v) - T_{cont}} \right)
\end{equation}
where T$_{on-off}$ is the difference between the HISA (T$_{on}$) and background (T$_{off}$) brightness, T$_{HISA}$ is the HISA temperature, $p$ a dimensionless parameter in the range [0-1] which accounts for foreground emission, and T$_{cont}$ the continuum brightness (which we ignore here). We here use $p$ = 1 which assumes no foreground emission for T$_{off}$. From this it is then possible to calculate the HISA column density using \citep{Wilson2009}
\begin{equation}
    N_{HISA}\,[cm^{-2}] = 1.8224\times 10^{18}\,\, T_{HISA}\,[K] \int \tau_{HISA}(v)dv\,[km\,s^{-1}]
\end{equation}
Here, we are however hampered by the noisy spectra. Therefore, we assume a constant $\tau_{HISA}$ with T$_{on-off}$ = -137 K and T$_{off}$ = 159 K. T$_{on-off}$ and T$_{off}$ were calculated based on the central and peak intensity of the ring mentioned above with an assumed FWHM of 2 km s$^{-1}$. Assuming T$_{HISA}$ = 20 K this gives rise to a required N$_{HISA}$ values of 3$\times$10$^{20}$ cm$^{-2}$. For T$_{HISA}$ = 10 K this drops to 9$\times$10$^{19}$ cm$^{-2}$, but at T$_{HISA}$ = 30 K no solution can be found. For HISA temperatures of only 20 K or less it is thus possible to find a solution. In addition, the column density associated with HISA is well below the total column density at the tip of IC 63 (A$_{V}$ $\approx$ 2 \citep{vanDePutte2020}, accounting for fore- and background extinction). With an A$_{V} \approx$ 2, or N$_{H} \approx$ 4.4$\times$10$^{21}$ cm$^{-3}$ (using \citet{Guver2009}), about 2-7\% of the gas in the line-of-sight would be associated to HISA. However, it has to be noted that these are significant column densities for such cold HISA \citep{Wang2020,Kabanovic2022,Seifried2022}. In addition, for a path length of 0.05 pc the typical density would be of the order of 0.6-3$\times$10$^{3}$ cm$^{-3}$ which is relatively high. It is thus possible to explain the ring morphology in IC~63 with \HI\ self-absorption, but the HISA conditions would have to be quite extreme compared to typical values in the Galactic ISM. More sensitive observations will be able to properly address this question.\\\\
Another option to explain the ring as well as the clumpy structure of \HI\ in the PDR would be a chemical effect that removes the presence of a layered PDR. Work by \citet{Stoerzer1998} proposed a criterion $\chi$/n $<$ 0.2v$_{IF}$
\footnote{With $\chi$ the FUV field strength \citep{Draine1978}, n the density and v$_{IF}$ the ionization front (IF) velocity.} where a PDR is dominated by non-equilibrium chemistry that removes the presence of a \HI/H$_{2}$ transition front. The recent semi-analytical solution by \citet{Maillard2021} proposes a more involved criterion for the transition to non-equilibrium PDR chemistry and no clearly defined \HI/H$_{2}$ transition front (see their Eq. 16). To examine the potential role of non-equilibrium chemistry in the IC 63 PDR, we compute the typical ionization front velocity (v$_{IF}$) for the head of this region using
\begin{equation}
    \text{v}_{IF} = \frac{\text{n}_{e}}{\text{n}_{PDR}}\text{v}_{II}
\end{equation}
with n$_{e}$ the electron density on the \HII\ side of the ionization front, a velocity v$_{II} \approx$ c$_{II} \approx$ 10 km s$^{-1}$ \citep[i.e. the sound speed of the ionized gas;][]{Stoerzer1998} and n$_{PDR}$ the density of the PDR. 
To calculate the electron density (n$_{e}$), we use the 1.4 GHz NRAO VLA Sky Survey (NVSS) radio continuum data \citep{Condon1998}. At the tip of IC 63, we typically find values of 0.05 Jy beam$^{-1}$ with a beam size of 45$^{\prime\prime}$. Since the radio continuum spectral index analysis from \citet{Blouin1997} indicates that the emission at 1.4 GHz is optically thin in IC 63, we can use
\begin{equation}
    {\scriptstyle \left( \frac{\text{EM}}{\text{pc cm}^{-6}}\right) = 3.217\times10^{7}\left( \frac{F_{\nu}}{Jy}\right)\left( \frac{\nu}{GHz}\right)^{0.1}\left( \frac{T_{e}}{K}\right)^{0.35}\left( \frac{\theta_{\text{source}}}{\text{arcsec}}\right)^{-2}}
\end{equation}
taken from \citet{Schmiedeke2016} to calculate the emission measure (EM). Here, F$_{\nu}$ (= 0.05 Jy) is the flux density, T$_{e}$ (= 8000 K) the electron temperature, $\nu$ (= 1.4 GHz) the observed frequency and $\theta_{\text{source}}$ (= 45$^{\prime\prime}$) the studied aperture. This results in EM = 1.9$\times$10$^{3}$ pc cm$^{-6}$. Using a size of 0.05 pc results in n$_{e}$ = 2.0$\times$10$^{2}$ cm$^{-3}$. To calculate the density in the PDR, \citet{vanDePutte2020} found an A$_{V}$ $\approx$ 2 for IC 63. Combining this with a typical size of 0.05 pc for the clumps gives a proton density n$_{H}$ = 2.8$\times$10$^{4}$ cm$^{-3}$ using the relation from \citet{Guver2009} that links A$_{V}$ to column density. Combining this gives v$_{IF}$ = 7.0$\times$10$^{-2}$ km s$^{-1}$. To determine whether the PDR is governed by non-equilibrium processes we have to estimate the FUV field strength. For the FUV field strength, the original literature proposes a value of $\chi \approx$ 6$\times$10$^{2}$ \citep{Jansen1994} at a distance of 1.3 pc from the ionizing star $\gamma$ Cas. We verified this value, using a distance of 1.3 to 2 pc from $\gamma$ Cas (Caputo et al. 2023, submitted) with the method presented in \citet{Bonne2020}. This directly calculates the FUV field strength at a specific distance from an OB star (or cluster) using the emitted radiation as a function of wavelength from the OB stars in the \citet{Kurucz1979} models. This results in G$_{0}$ = 1.1-2.6$\times$10$^{2}$ (with G$_{0}$ = 1.7$\chi$, this leads to $\chi$ = 0.65-1.5$\times$10$^{2}$) which is significantly lower than the prediction by \citet{Jansen1994}. However, the value calculated here is consistent with more recent work by \citet{Andrews2018} who proposed an FUV field strength $\chi$ $\approx$ 90 and \citet{2021ApJ...923..107S} who found $\chi$ = 0.6-2.4$\times$10$^{2}$. To investigate this in more detail, we use the \CII\ intensity presented in Caputo et al. (2023, submitted), which typically is 2.5-3.0$\times$10$^{-4}$ erg s$^{-1}$ cm$^{-2}$ sr$^{-1}$, and the PDR Toolbox \citep{Pound2008,Pound2023}. At densities of 1-5$\times$10$^{4}$ cm$^{-3}$ this \CII\ intensity corresponds to a typical FUV field of $\chi$ = 0.6-2.0$\times$10$^{2}$. This is consistent with the estimate based on the method in \citet{Bonne2020}. Using the calculated v$_{IF}$, the prescription by \citet{Stoerzer1998} provides a critical ($\chi$/n)$_{crit}$ = 1.4$\times$10$^{-2}$ and the prescription by \citet{Maillard2021} gives ($\chi$/n)$_{crit}$ = 2.4$\times$10$^{-3}$. Using the determined $\chi$ = 0.6-2.0$\times$10$^{2}$ range from the \CII\ observations gives $\chi$/n = 2.1-7.1$\times$10$^{-3}$ for IC 63. The results based on both criteria provide different conclusions. The \citet{Maillard2021} criterion suggests there likely still is a \HI/H$_{2}$ transition front whereas the \citet{Stoerzer1998} criterion suggests IC 63 no longer has a \HI/H$_{2}$ transition front which might provide an explanation for the more complex or clumpy \HI\ structure. However, it has to be noted that the \citet{Maillard2021} solution should be the more accurate one as it includes the dynamics induced by the photo-evaporation and dust shielding \citep{Sternberg2014}.\\
A last thing to consider is the dynamics in the PDR itself which might be significant when considering the apparent 1~km~s$^{-1}$ offset in velocity between \HI\ and \CII, in particular towards the tip of the region, in Fig. \ref{fig:velDiffMap}. In the previous section we obtained a dynamical timescale of 5$\times$10$^{4}$ yr for the PDR which we can compare with the H$_{2}$ dissociation timescale. Based on \citet{Draine1996} and \citet{Stoerzer1998}, this is given by
\begin{equation}
    \text{t}_{H_{2}} \approx 0.6\frac{\text{N(H}_{2}\text{)}^{3/4}}{\chi} \text{s}
\end{equation}
Using N(H$_{2}$) = $0.5\times10^{21}$ cm$^{-2}$ \citep{Stoerzer1998} and $\chi$ = 0.6-2.0$\times$10$^{2}$ results in a dissociation timescale of 0.3-1.0 Myr. It thus appears that the dynamical timescale is significantly shorter than the H$_{2}$ dissociation timescale. As a result the dynamics in the PDR can reorganize the internal structure and so remove a plane-parallel \HI/H$_{2}$ structure which could explain the observed clumpy \HI\ structure.

\section{Conclusions} \label{concl}
We have presented combined interferometric \HI\ observations from the GMRT, WSRT and DRAO of IC 63 with an effective angular resolution of 15$^{\prime\prime}$ which allows to resolve the structure of the PDR in IC 63. These observations are further complemented with SOFIA \CII\ data. To properly analyze the \HI\ observations at the low Galactic latitude of IC~63 (b=-2), we subtract the galactic background from the spectrum using nearby regions devoid of \HI\ emission associated with IC 63. Although the S/N is somewhat limited we conclude that the \HI\ structure of IC 63 appears to be clumpy, and that the tip of IC 63 appears to have a ring-like morphology in \HI. With analytical models, we show that this ring-like structure is likely not the projection of an unfilled spherical shell. However, more sensitive observations with new and future observatories can help to confirm these results and explore them in further detail. We also obtain a view on the dynamics. This shows dynamic structure with $\Delta$v $\approx$ 1~km~s$^{-1}$ inside the tip of the PDR. In addition, the full region consists of several legs that appear to be moving apart. This suggests that IC 63 will completely disperse within less than 0.5 Myr.\\
Analyzing the \HI\ spectra and PDR characteristics, we conclude that the dynamical timescale in the PDR is too short for there to be a plane-parallel \HI/H$_{2}$ structure which could explain the clumpy \HI\ structure. The S/N is not sufficient to conclude on the contribution of HISA to the spectra, but if the HISA temperature is below 20 K then HISA can contribute to the apparent clumpy \HI\ structure. We also investigated whether the ionization front velocity leading to non-equilibrium chemistry could explain the clumpy structure. Different prescriptions for the transition to non-equilibrium chemistry due to the ionization front velocity lead to different conclusions, but the most recent and accurate prescription by \citet{Maillard2021} concludes that non-equilibrium chemistry should not play a major role. We thus tentatively propose that the apparent clumpy \HI\ PDR structure is predominantly explained by the short dynamical timescale in the PDR, but it cannot be excluded that different mechanisms contribute to this observed clumpy \HI\ structure.

\acknowledgments
We thank the anonymous referee for detailed comments that helped to improve the clarity of the paper significantly. We thank G. Barentsen for fruitful exchange on the use of the IPHAS dataset. We acknowledge the professional and patient support of the WSRT staff, in particular Gyula Jozsa and Adriaan Renting. We thank the staff of the GMRT that made these observations possible. GMRT is run by the National Centre for Radio Astrophysics of the Tata Institute of Fundamental Research. This research used the facilities of the Canadian Astronomy Data Centre operated by the National Research Council of Canada with the support of the Canadian Space Agency. L.B. was supported by a USRA postdoctoral fellowship, funded through the NASA SOFIA contract NNA17BF53C. B-G A., A.S. and J.K. gratefully acknowledge the support from the National Science Foundation (NSF) under grant AST-1715876 to USRA.  K.K. and J.Y., gratefully acknowledge the support from NSF under grant AST-1715060 to SCU. J.Y. was supported by a Fox Fellowship from Santa Clara University. R.M. acknowledges support from the National Radio Astronomy Observatory (NRAO). The National Radio Astronomy Observatory is a facility of the National Science Foundation operated under cooperative agreement by Associated Universities, Inc.  Based, in part, on observations made with the NASA/DLR Stratospheric Observatory for Infrared Astronomy (SOFIA). SOFIA is jointly operated by the Universities Space Research Association, Inc. (USRA), under NASA contract NAS2-97001, and the Deutsches SOFIA Institut (DSI) under DLR contract 50 OK 0901 to the University of Stuttgart.

\vspace{5mm}
\facilities{GMRT, WSRT, DRAO, SOFIA(upGREAT), CFHT(WIRCam), CADC}


\software{astropy \citep{2013A&A...558A..33A}, MIRIAD \citep{1995ASPC...77..433S}, AIPS \citep{2003ASSL..285..109G}}



\appendix

\section{CLEAN vs. MEM reduction}\label{sec:CLEANvsMEM}
\begin{figure*}
    \centering
    \includegraphics[width=0.495\hsize]{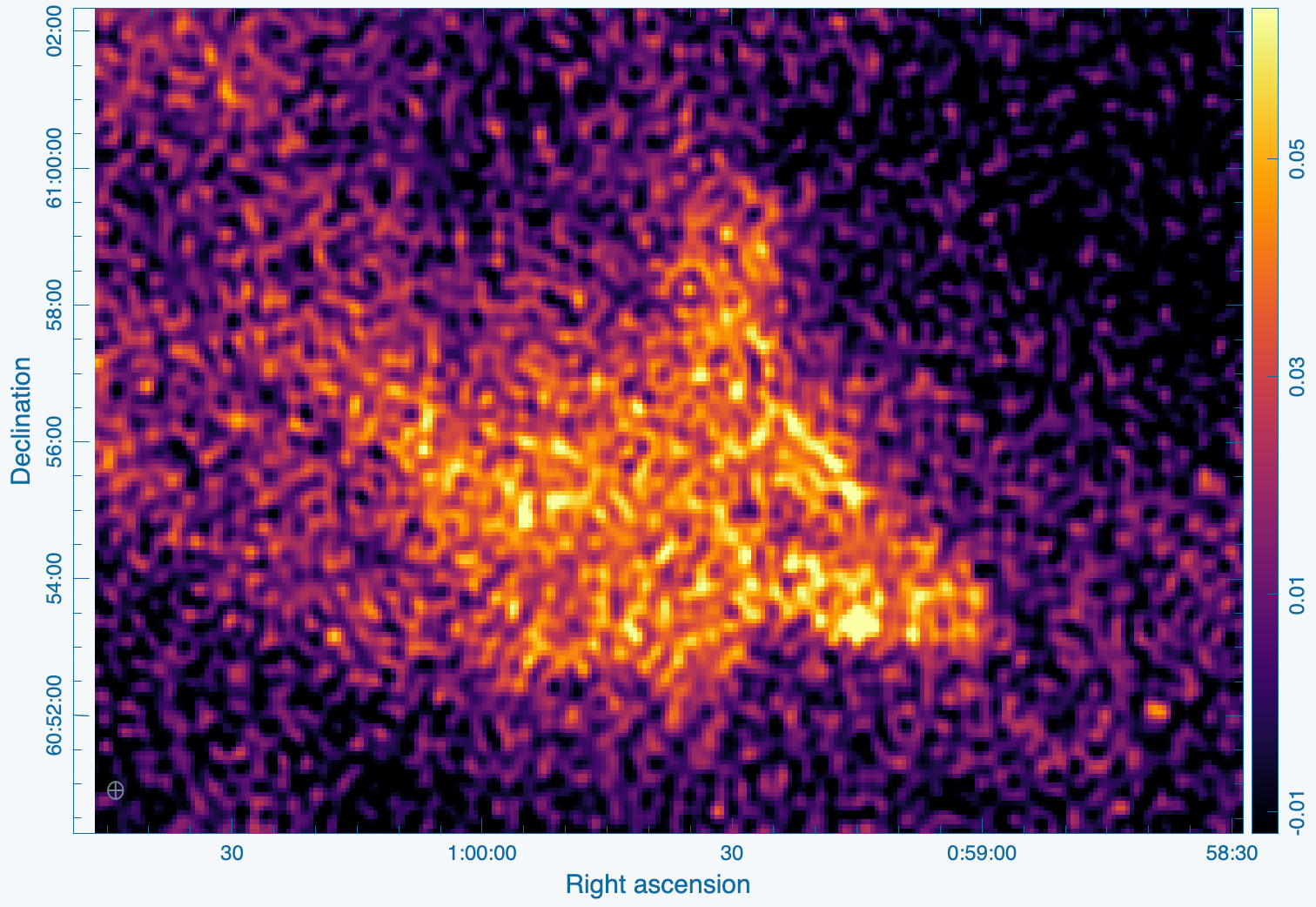}
    \includegraphics[width=0.495\hsize]{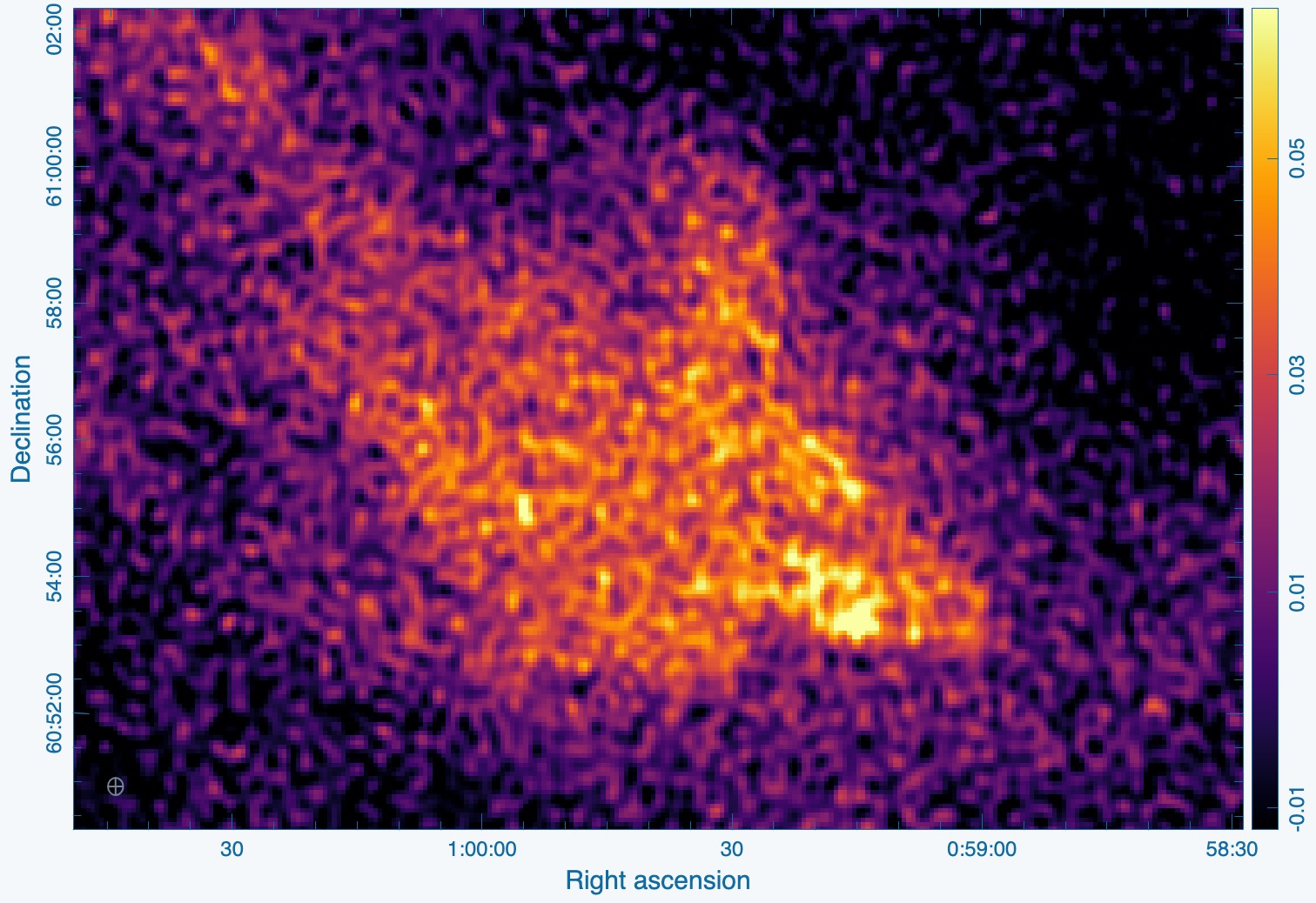}
    \caption{{\bf Left}:  The moment zero map for IC 63 between -2.5 and 4.5 km s$^{-1}$ obtained using the MEM reduction technique. {\bf Right}: The same using the CLEAN reduction technique, showing a similar clumpy morphology.}
    \label{fig:MEMvsCLEAN}
\end{figure*}

Figure \ref{fig:MEMvsCLEAN} displays the moment zero map for the \HI\ emission between -2.5 and 4.5 km s$^{-1}$, which is not background subtracted, when reduced using the MEM technique (left) and the CLEAN technique (right). Both methods create a very similar map which indicates that the clumpy morphology for the region is not an artefact. 

\section{The individual \HI\ profiles and spectrum over the ring}\label{sec:indHIprofiles}
\begin{figure}
    \centering
    \raisebox{-0.5\height}{\includegraphics[width=0.37\hsize]{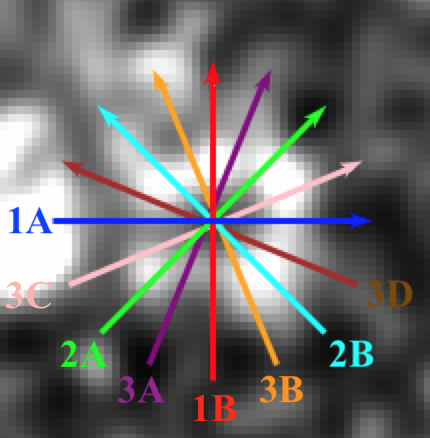}}
    \hspace{0.2in}
    \raisebox{-0.5\height}{\includegraphics[width=0.58\hsize]{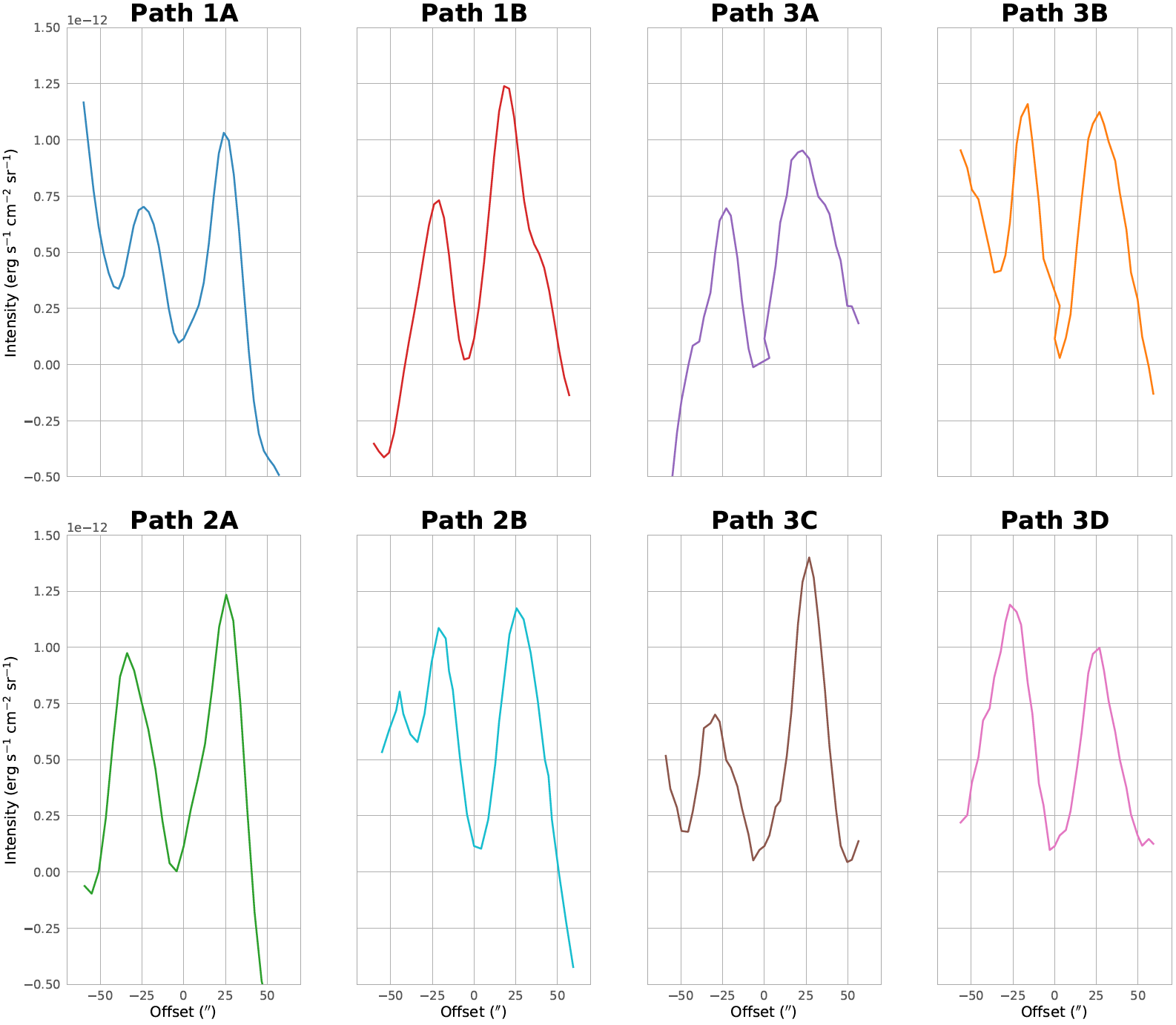}}
    \caption{{\bf Left}: The different paths that were defined over the \HI\ ring-like feature on top of the \HI\ integrated intensity map. {\bf Right}: The \HI\ emission profiles over the ring for the 8 individual paths defined on the left.}
    \label{fig:HIprofiles}
\end{figure}

\begin{figure}
\includegraphics[width=0.33\textwidth]{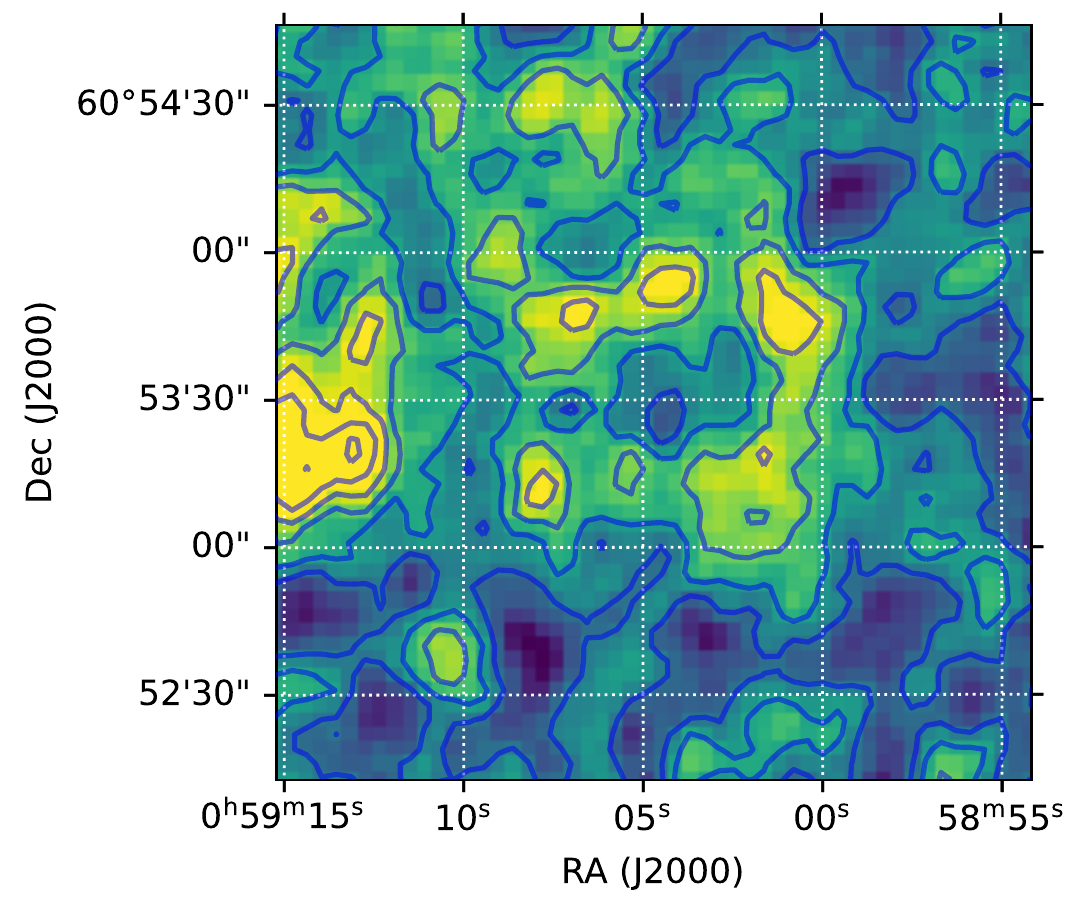}
\includegraphics[width=0.33\textwidth]{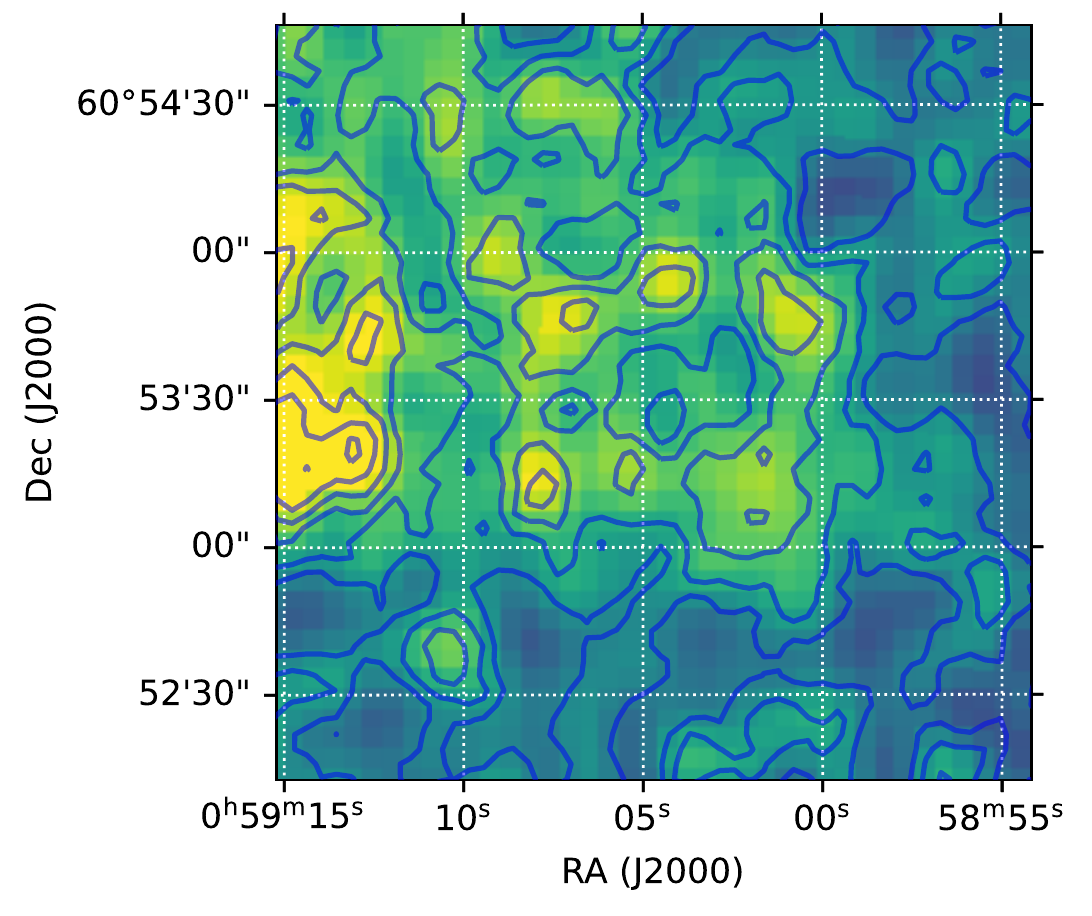}
\includegraphics[width=0.33\textwidth]{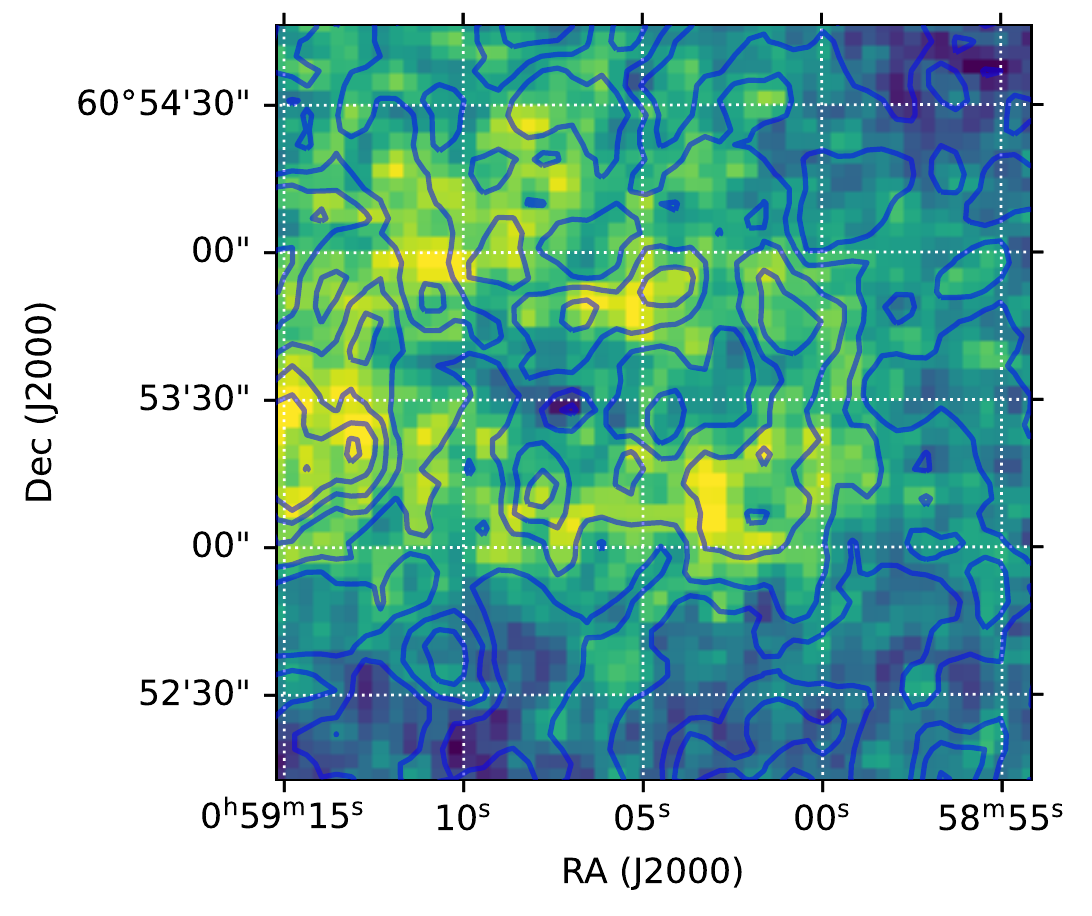}
\caption{\textbf{Left:} The ring seen in the moment 0 map from the combined cube with contours from 0.1 to 0.225 Jy\,km\,s$^{-1}$ in steps of 0.025 Jy\,km\,s$^{-1}$. \textbf{Center:} The same field in the moment 0 map from the WSRT data with the contours from the combined cube overlaid. \textbf{Right:} The same field in the moment 0 map from the GMRT data with the contours from the combined cube overlaid. These plots show that the ring is seen in both interferometric datasets and so is not an artefact from a single telescope that has propagated into our combined cube. They also reveal that there appears to be some substructure in the ring.
}
\label{fig:ring}
\end{figure}

\begin{figure}
    \centering
    \raisebox{-0.5\height}{\includegraphics[width=0.5\hsize]{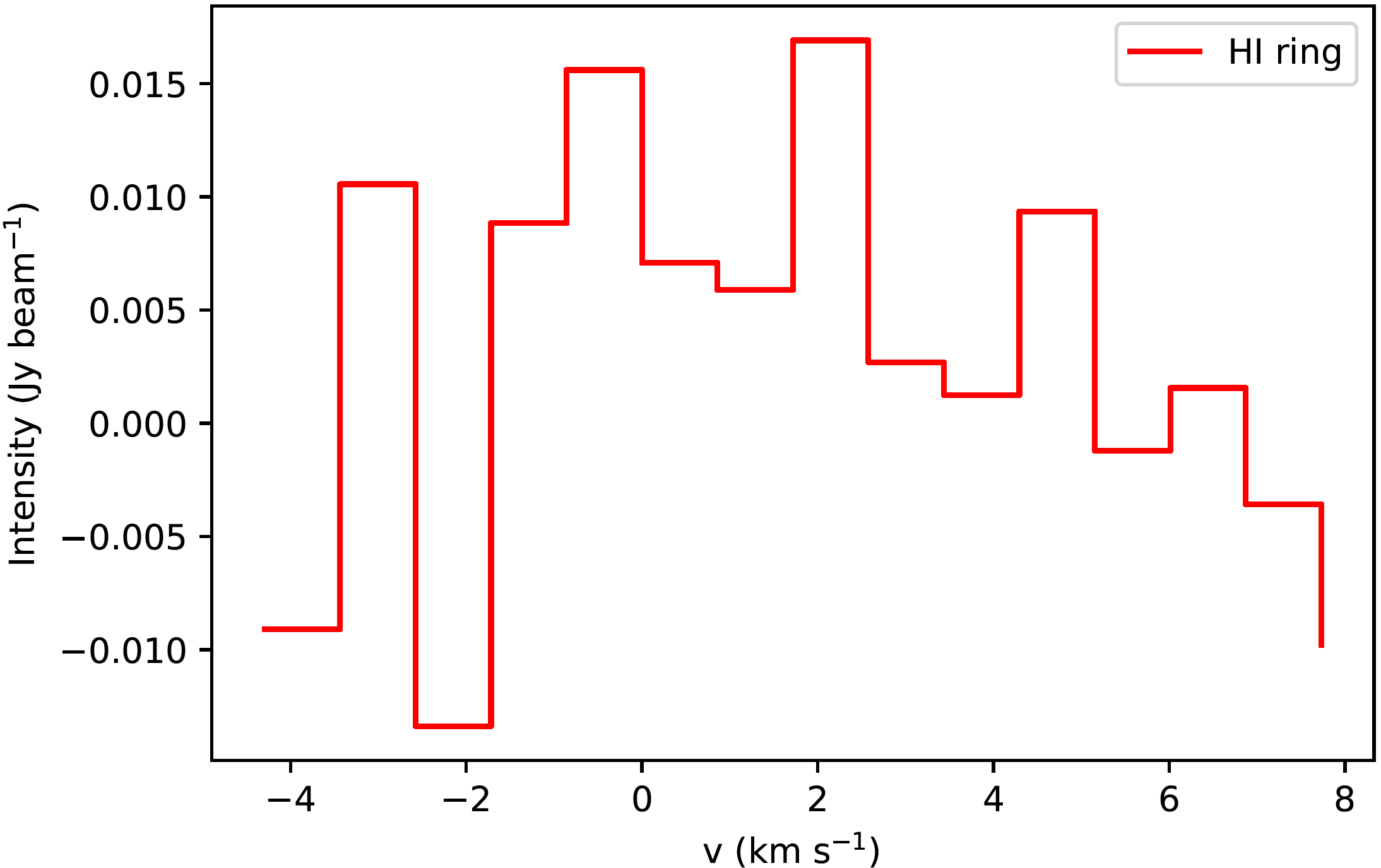}}
    \caption{The average \HI\ spectrum obtained from inside the ring at the tip of IC 63. As expected the noise is fairly high, making it challenging to conclude on the presence of self-absorption.}
    \label{fig:HIspecRing}
\end{figure}

To obtain a better view of the ring-like feature, we created 8 cuts through the ring with in steps of 20$^{o}$, which is shown in Fig. \ref{fig:HIprofiles}. The resulting profiles along the cut are presented in Fig. \ref{fig:HIprofiles}. The center was defined by making sure that the offset from the center to the maximum intensity is equal in the positive and negative direction along the vertical and horizontal axes. Within a typical offset of $\pm$ 40$^{\prime\prime}$, the profiles show similar behaviour with peak values varying between 0.7$\times$10$^{-12}$ and 1.4$\times$10$^{-12}$ erg s$^{-1}$ cm$^{-2}$ sr$^{-1}$ while the minimal value towards the center of the ring is relatively close to 1.0$\times$10$^{-13}$ erg s$^{-1}$ cm$^{-2}$ sr$^{-1}$.\\
To further verify that the ring-like feature at the tip is not an image artefact, the moment 0 map was created for the WSRT and GMRT maps individually which both show a dip towards the same location that is seen in the full combined map, see Fig. \ref{fig:ring}. This gives further confidence that this drop in emission towards the center of the tip is a real feature. Additionally, Fig. \ref{fig:ring} also indicates that the weaker emission in the eastern part of the ring might be attributed to an artefact in the GMRT data cube.\\
Lastly, we also present the average spectrum with the best possible S/N extracted from inside the ring, see Fig. \ref{fig:HIspecRing}. This spectrum shows a non-Gaussian skewed line profile, but because of the relatively high noise rms in the data it is not possible to assess whether this non-Gaussian profile is due to self-absorption or not.

\bibliography{IC63_HI}{}

\begin{thebibliography}{}
\expandafter\ifx\csname natexlab\endcsname\relax\def\natexlab#1{#1}\fi
\providecommand{\url}[1]{\href{#1}{#1}}
\providecommand{\dodoi}[1]{doi:~\href{http://doi.org/#1}{\nolinkurl{#1}}}
\providecommand{\doeprint}[1]{\href{http://ascl.net/#1}{\nolinkurl{http://ascl.net/#1}}}
\providecommand{\doarXiv}[1]{\href{https://arxiv.org/abs/#1}{\nolinkurl{https://arxiv.org/abs/#1}}}

\bibitem[{{Andersson} {et~al.}(2013){Andersson}, {Piirola}, {De Buizer},
  {Clemens}, {Uomoto}, {Charcos-Llorens}, {Geballe}, {Lazarian}, {Hoang}, \&
  {Vornanen}}]{Andersson2013}
{Andersson}, B.~G., {Piirola}, V., {De Buizer}, J., {et~al.} 2013, \apj, 775,
  84, \dodoi{10.1088/0004-637X/775/2/84}

\bibitem[{{Andrews} {et~al.}(2018){Andrews}, {Peeters}, {Tielens}, \&
  {Okada}}]{Andrews2018}
{Andrews}, H., {Peeters}, E., {Tielens}, A.~G.~G.~M., \& {Okada}, Y. 2018,
  \aap, 619, A170, \dodoi{10.1051/0004-6361/201832808}

\bibitem[{{Astropy Collaboration} {et~al.}(2013){Astropy Collaboration},
  {Robitaille}, {Tollerud}, {Greenfield}, {Droettboom}, {Bray}, {Aldcroft},
  {Davis}, {Ginsburg}, {Price-Whelan}, {Kerzendorf}, {Conley}, {Crighton},
  {Barbary}, {Muna}, {Ferguson}, {Grollier}, {Parikh}, {Nair}, {Unther},
  {Deil}, {Woillez}, {Conseil}, {Kramer}, {Turner}, {Singer}, {Fox}, {Weaver},
  {Zabalza}, {Edwards}, {Azalee Bostroem}, {Burke}, {Casey}, {Crawford},
  {Dencheva}, {Ely}, {Jenness}, {Labrie}, {Lim}, {Pierfederici}, {Pontzen},
  {Ptak}, {Refsdal}, {Servillat}, \& {Streicher}}]{2013A&A...558A..33A}
{Astropy Collaboration}, {Robitaille}, T.~P., {Tollerud}, E.~J., {et~al.} 2013,
  \aap, 558, A33, \dodoi{10.1051/0004-6361/201322068}

\bibitem[{{Bertoldi} \& {Draine}(1996)}]{Bertoldi1996}
{Bertoldi}, F., \& {Draine}, B.~T. 1996, \apj, 458, 222, \dodoi{10.1086/176805}

\bibitem[{{Bisbas} {et~al.}(2012){Bisbas}, {Bell}, {Viti}, {Yates}, \&
  {Barlow}}]{Bisbas2012}
{Bisbas}, T.~G., {Bell}, T.~A., {Viti}, S., {Yates}, J., \& {Barlow}, M.~J.
  2012, \mnras, 427, 2100, \dodoi{10.1111/j.1365-2966.2012.22077.x}

\bibitem[{{Blouin} {et~al.}(1997){Blouin}, {McCutcheon}, {Dewdney}, {Roger},
  {Purton}, {Kester}, \& {Bontekoe}}]{Blouin1997}
{Blouin}, D., {McCutcheon}, W.~H., {Dewdney}, P.~E., {et~al.} 1997, \mnras,
  287, 455, \dodoi{10.1093/mnras/287.2.455}

\bibitem[{{Bonne} {et~al.}(2020){Bonne}, {Schneider}, {Bontemps}, {Clarke},
  {Gusdorf}, {Lehmann}, {Steinke}, {Csengeri}, {Kabanovic}, {Simon},
  {Buchbender}, \& {G{\"u}sten}}]{Bonne2020}
{Bonne}, L., {Schneider}, N., {Bontemps}, S., {et~al.} 2020, \aap, 641, A17,
  \dodoi{10.1051/0004-6361/201937104}

\bibitem[{{Bonne} {et~al.}(2022){Bonne}, {Schneider}, {Garc{\'\i}a}, {Bij},
  {Broos}, {Fissel}, {Guesten}, {Jackson}, {Simon}, {Townsley}, {Zavagno},
  {Aladro}, {Buchbender}, {Guevara}, {Higgins}, {Jacob}, {Kabanovic}, {Karim},
  {Soam}, {Stutzki}, {Tiwari}, {Wyrowski}, \& {Tielens}}]{Bonne2022}
{Bonne}, L., {Schneider}, N., {Garc{\'\i}a}, P., {et~al.} 2022, \apj, 935, 171,
  \dodoi{10.3847/1538-4357/ac8052}

\bibitem[{{Condon} {et~al.}(1998){Condon}, {Cotton}, {Greisen}, {Yin},
  {Perley}, {Taylor}, \& {Broderick}}]{Condon1998}
{Condon}, J.~J., {Cotton}, W.~D., {Greisen}, E.~W., {et~al.} 1998, \aj, 115,
  1693, \dodoi{10.1086/300337}

\bibitem[{{Draine}(1978)}]{Draine1978}
{Draine}, B.~T. 1978, \apjs, 36, 595, \dodoi{10.1086/190513}

\bibitem[{{Draine} \& {Bertoldi}(1996)}]{Draine1996}
{Draine}, B.~T., \& {Bertoldi}, F. 1996, \apj, 468, 269, \dodoi{10.1086/177689}

\bibitem[{{Fleming} {et~al.}(2010){Fleming}, {France}, {Lupu}, \&
  {McCandliss}}]{Fleming2010}
{Fleming}, B., {France}, K., {Lupu}, R.~E., \& {McCandliss}, S.~R. 2010, \apj,
  725, 159, \dodoi{10.1088/0004-637X/725/1/159}

\bibitem[{{France} {et~al.}(2005){France}, {Andersson}, {McCandliss}, \&
  {Feldman}}]{France2005}
{France}, K., {Andersson}, B.~G., {McCandliss}, S.~R., \& {Feldman}, P.~D.
  2005, \apj, 628, 750, \dodoi{10.1086/430878}

\bibitem[{{Gierens} {et~al.}(1992){Gierens}, {Stutzki}, \&
  {Winnewisser}}]{Gierens1992}
{Gierens}, K.~M., {Stutzki}, J., \& {Winnewisser}, G. 1992, \aap, 259, 271

\bibitem[{{Greisen}(2003)}]{2003ASSL..285..109G}
{Greisen}, E.~W. 2003, {AIPS, the VLA, and the VLBA}, ed. A.~{Heck}, Vol. 285,
  109, \dodoi{10.1007/0-306-48080-8\_7}

\bibitem[{{Guan} {et~al.}(2012){Guan}, {Stutzki}, {Graf}, {G{\"u}sten},
  {Okada}, {Requena-Torres}, {Simon}, \& {Wiesemeyer}}]{Guan2012}
{Guan}, X., {Stutzki}, J., {Graf}, U.~U., {et~al.} 2012, \aap, 542, L4,
  \dodoi{10.1051/0004-6361/201218925}

\bibitem[{{G{\"u}ver} \& {{\"O}zel}(2009)}]{Guver2009}
{G{\"u}ver}, T., \& {{\"O}zel}, F. 2009, \mnras, 400, 2050,
  \dodoi{10.1111/j.1365-2966.2009.15598.x}

\bibitem[{{Heyminck} {et~al.}(2012){Heyminck}, {Graf}, {G{\"u}sten}, {Stutzki},
  {H{\"u}bers}, \& {Hartogh}}]{Heyminck2012}
{Heyminck}, S., {Graf}, U.~U., {G{\"u}sten}, R., {et~al.} 2012, \aap, 542, L1,
  \dodoi{10.1051/0004-6361/201218811}

\bibitem[{{Hollenbach} \& {Tielens}(1999)}]{Hollenbach1999}
{Hollenbach}, D.~J., \& {Tielens}, A.~G.~G.~M. 1999, Reviews of Modern Physics,
  71, 173, \dodoi{10.1103/RevModPhys.71.173}

\bibitem[{{Jansen} {et~al.}(1994){Jansen}, {van Dishoeck}, \&
  {Black}}]{Jansen1994}
{Jansen}, D.~J., {van Dishoeck}, E.~F., \& {Black}, J.~H. 1994, \aap, 282, 605

\bibitem[{{Kabanovic} {et~al.}(2022){Kabanovic}, {Schneider},
  {Ossenkopf-Okada}, {Falasca}, {G{\"u}sten}, {Stutzki}, {Simon}, {Buchbender},
  {Anderson}, {Bonne}, {Guevara}, {Higgins}, {Koribalski}, {Luisi}, {Mertens},
  {Okada}, {R{\"o}llig}, {Seifried}, {Tiwari}, {Wyrowski}, {Zavagno}, \&
  {Tielens}}]{Kabanovic2022}
{Kabanovic}, S., {Schneider}, N., {Ossenkopf-Okada}, V., {et~al.} 2022, \aap,
  659, A36, \dodoi{10.1051/0004-6361/202142575}

\bibitem[{{Karr} {et~al.}(2005){Karr}, {Noriega-Crespo}, \&
  {Martin}}]{Karr2005}
{Karr}, J.~L., {Noriega-Crespo}, A., \& {Martin}, P.~G. 2005, \aj, 129, 954,
  \dodoi{10.1086/426912}

\bibitem[{{Kaufman} {et~al.}(1999){Kaufman}, {Wolfire}, {Hollenbach}, \&
  {Luhman}}]{Kaufman1999}
{Kaufman}, M.~J., {Wolfire}, M.~G., {Hollenbach}, D.~J., \& {Luhman}, M.~L.
  1999, \apj, 527, 795, \dodoi{10.1086/308102}

\bibitem[{{Kurucz}(1979)}]{Kurucz1979}
{Kurucz}, R.~L. 1979, \apjs, 40, 1, \dodoi{10.1086/190589}

\bibitem[{{Landecker} {et~al.}(2000){Landecker}, {Dewdney}, {Burgess}, {Gray},
  {Higgs}, {Hoffmann}, {Hovey}, {Karpa}, {Lacey}, {Prowse}, {Purton}, {Roger},
  {Willis}, {Wyslouzil}, {Routledge}, \& {Vaneldik}}]{Landecker2000}
{Landecker}, T.~L., {Dewdney}, P.~E., {Burgess}, T.~A., {et~al.} 2000, \aaps,
  145, 509, \dodoi{10.1051/aas:200025710.48550/arXiv.astro-ph/0006415}

\bibitem[{{Le Petit} {et~al.}(2006){Le Petit}, {Nehm{\'e}}, {Le Bourlot}, \&
  {Roueff}}]{LePetit2006}
{Le Petit}, F., {Nehm{\'e}}, C., {Le Bourlot}, J., \& {Roueff}, E. 2006, \apjs,
  164, 506, \dodoi{10.1086/503252}

\bibitem[{{Maillard} {et~al.}(2021){Maillard}, {Bron}, \& {Le
  Petit}}]{Maillard2021}
{Maillard}, V., {Bron}, E., \& {Le Petit}, F. 2021, \aap, 656, A65,
  \dodoi{10.1051/0004-6361/202140865}

\bibitem[{{McClure-Griffiths} {et~al.}(2005){McClure-Griffiths}, {Dickey},
  {Gaensler}, {Green}, {Haverkorn}, \& {Strasser}}]{McClureGriffiths2005}
{McClure-Griffiths}, N.~M., {Dickey}, J.~M., {Gaensler}, B.~M., {et~al.} 2005,
  \apjs, 158, 178, \dodoi{10.1086/430114}

\bibitem[{{Pound} \& {Wolfire}(2008)}]{Pound2008}
{Pound}, M.~W., \& {Wolfire}, M.~G. 2008, in Astronomical Society of the
  Pacific Conference Series, Vol. 394, Astronomical Data Analysis Software and
  Systems XVII, ed. R.~W. {Argyle}, P.~S. {Bunclark}, \& J.~R. {Lewis}, 654

\bibitem[{{Pound} \& {Wolfire}(2023)}]{Pound2023}
{Pound}, M.~W., \& {Wolfire}, M.~G. 2023, \aj, 165, 25,
  \dodoi{10.3847/1538-3881/ac9b1f}

\bibitem[{{Risacher} {et~al.}(2016){Risacher}, {G{\"u}sten}, {Stutzki},
  {H{\"u}bers}, {Bell}, {Buchbender}, {B{\"u}chel}, {Csengeri}, {Graf},
  {Heyminck}, {Higgins}, {Honingh}, {Jacobs}, {Klein}, {Okada}, {Parikka},
  {P{\"u}tz}, {Reyes}, {Ricken}, {Riquelme}, {Simon}, \&
  {Wiesemeyer}}]{Risacher2016}
{Risacher}, C., {G{\"u}sten}, R., {Stutzki}, J., {et~al.} 2016, \aap, 595, A34,
  \dodoi{10.1051/0004-6361/201629045}

\bibitem[{{Risacher} {et~al.}(2018){Risacher}, {G{\"u}sten}, {Stutzki},
  {H{\"u}bers}, {Aladro}, {Bell}, {Buchbender}, {B{\"u}chel}, {Csengeri},
  {Duran}, {Graf}, {Higgins}, {Honingh}, {Jacobs}, {Justen}, {Klein},
  {Mertens}, {Okada}, {Parikka}, {P{\"u}tz}, {Reyes}, {Richter}, {Ricken},
  {Riquelme}, {Rothbart}, {Schneider}, {Simon}, {Wienold}, {Wiesemeyer},
  {Ziebart}, {Fusco}, {Rosner}, \& {Wohler}}]{Risacher2018}
---. 2018, Journal of Astronomical Instrumentation, 7, 1840014,
  \dodoi{10.1142/S2251171718400147}

\bibitem[{{R{\"o}llig} {et~al.}(2007){R{\"o}llig}, {Abel}, {Bell}, {Bensch},
  {Black}, {Ferland}, {Jonkheid}, {Kamp}, {Kaufman}, {Le Bourlot}, {Le Petit},
  {Meijerink}, {Morata}, {Ossenkopf}, {Roueff}, {Shaw}, {Spaans}, {Sternberg},
  {Stutzki}, {Thi}, {van Dishoeck}, {van Hoof}, {Viti}, \&
  {Wolfire}}]{Roellig2007}
{R{\"o}llig}, M., {Abel}, N.~P., {Bell}, T., {et~al.} 2007, \aap, 467, 187,
  \dodoi{10.1051/0004-6361:20065918}

\bibitem[{{Roy} {et~al.}(2010){Roy}, {Gupta}, {Pen}, {Peterson}, {Kudale}, \&
  {Kodilkar}}]{Roy2010}
{Roy}, J., {Gupta}, Y., {Pen}, U.-L., {et~al.} 2010, Experimental Astronomy,
  28, 25, \dodoi{10.1007/s10686-010-9187-0}

\bibitem[{{Sault} {et~al.}(1995){Sault}, {Teuben}, \&
  {Wright}}]{1995ASPC...77..433S}
{Sault}, R.~J., {Teuben}, P.~J., \& {Wright}, M.~C.~H. 1995, in Astronomical
  Society of the Pacific Conference Series, Vol.~77, Astronomical Data Analysis
  Software and Systems IV, ed. R.~A. {Shaw}, H.~E. {Payne}, \& J.~J.~E.
  {Hayes}, 433.
\newblock \doarXiv{astro-ph/0612759}

\bibitem[{{Schmiedeke} {et~al.}(2016){Schmiedeke}, {Schilke}, {M{\"o}ller},
  {S{\'a}nchez-Monge}, {Bergin}, {Comito}, {Csengeri}, {Lis}, {Molinari},
  {Qin}, \& {Rolffs}}]{Schmiedeke2016}
{Schmiedeke}, A., {Schilke}, P., {M{\"o}ller}, T., {et~al.} 2016, \aap, 588,
  A143, \dodoi{10.1051/0004-6361/201527311}

\bibitem[{{Schneider} {et~al.}(2020){Schneider}, {Simon}, {Guevara},
  {Buchbender}, {Higgins}, {Okada}, {Stutzki}, {G{\"u}sten}, {Anderson},
  {Bally}, {Beuther}, {Bonne}, {Bontemps}, {Chambers}, {Csengeri}, {Graf},
  {Gusdorf}, {Jacobs}, {Justen}, {Kabanovic}, {Karim}, {Luisi}, {Menten},
  {Mertens}, {Mookerjea}, {Ossenkopf-Okada}, {Pabst}, {Pound}, {Richter},
  {Reyes}, {Ricken}, {R{\"o}llig}, {Russeil}, {S{\'a}nchez-Monge}, {Sandell},
  {Tiwari}, {Wiesemeyer}, {Wolfire}, {Wyrowski}, {Zavagno}, \&
  {Tielens}}]{Schneider2020}
{Schneider}, N., {Simon}, R., {Guevara}, C., {et~al.} 2020, \pasp, 132, 104301,
  \dodoi{10.1088/1538-3873/aba840}

\bibitem[{{Seifried} {et~al.}(2022){Seifried}, {Beuther}, {Walch}, {Syed},
  {Soler}, {Girichidis}, \& {W{\"u}nsch}}]{Seifried2022}
{Seifried}, D., {Beuther}, H., {Walch}, S., {et~al.} 2022, \mnras, 512, 4765,
  \dodoi{10.1093/mnras/stac607}

\bibitem[{{Soam} {et~al.}(2021{\natexlab{a}}){Soam}, {Andersson}, {Karoly},
  {DeWitt}, \& {Richter}}]{Soam2021c}
{Soam}, A., {Andersson}, B.~G., {Karoly}, J., {DeWitt}, C., \& {Richter}, M.
  2021{\natexlab{a}}, \apj, 923, 107, \dodoi{10.3847/1538-4357/ac2eb7}

\bibitem[{{Soam} {et~al.}(2021{\natexlab{b}}){Soam}, {Andersson}, {Karoly},
  {DeWitt}, \& {Richter}}]{2021ApJ...923..107S}
---. 2021{\natexlab{b}}, \apj, 923, 107,
  \dodoi{10.3847/1538-4357/ac2eb710.48550/arXiv.2110.11703}

\bibitem[{{Soam} {et~al.}(2017){Soam}, {Maheswar}, {Lee}, {Neha}, \&
  {Andersson}}]{soam2017}
{Soam}, A., {Maheswar}, G., {Lee}, C.~W., {Neha}, S., \& {Andersson}, B.~G.
  2017, \mnras, 465, 559, \dodoi{10.1093/mnras/stw2649}

\bibitem[{{Soam} {et~al.}(2021{\natexlab{c}}){Soam}, {Andersson},
  {Strai{\v{z}}ys}, {Caputo}, {Kazlauskas}, {Boyle}, {Janusz},
  {Zdanavi{\v{c}}ius}, \& {Acosta-Pulido}}]{soam2021b}
{Soam}, A., {Andersson}, B.~G., {Strai{\v{z}}ys}, V., {et~al.}
  2021{\natexlab{c}}, \aj, 161, 149, \dodoi{10.3847/1538-3881/abdd3b}

\bibitem[{{Sternberg} \& {Dalgarno}(1995)}]{Sternberg1995}
{Sternberg}, A., \& {Dalgarno}, A. 1995, \apjs, 99, 565, \dodoi{10.1086/192198}

\bibitem[{{Sternberg} {et~al.}(2014){Sternberg}, {Le Petit}, {Roueff}, \& {Le
  Bourlot}}]{Sternberg2014}
{Sternberg}, A., {Le Petit}, F., {Roueff}, E., \& {Le Bourlot}, J. 2014, \apj,
  790, 10, \dodoi{10.1088/0004-637X/790/1/10}

\bibitem[{{Stoerzer} {et~al.}(1996){Stoerzer}, {Stutzki}, \&
  {Sternberg}}]{Stoerzer1996}
{Stoerzer}, H., {Stutzki}, J., \& {Sternberg}, A. 1996, \aap, 310, 592

\bibitem[{{St{\"o}rzer} \& {Hollenbach}(1998)}]{Stoerzer1998}
{St{\"o}rzer}, H., \& {Hollenbach}, D. 1998, \apj, 495, 853,
  \dodoi{10.1086/305315}

\bibitem[{{Sugitani} {et~al.}(1991){Sugitani}, {Fukui}, \&
  {Ogura}}]{Sugitani1991}
{Sugitani}, K., {Fukui}, Y., \& {Ogura}, K. 1991, \apjs, 77, 59,
  \dodoi{10.1086/191597}

\bibitem[{{Taylor} {et~al.}(2017){Taylor}, {Leahy}, {Tian}, {Sunstrum},
  {Kothes}, {Landecker}, {Ransom}, \& {Higgs}}]{Taylor2017}
{Taylor}, A.~R., {Leahy}, D.~A., {Tian}, W.~W., {et~al.} 2017, \aj, 153, 113,
  \dodoi{10.3847/1538-3881/153/3/113}

\bibitem[{{Taylor} {et~al.}(2003){Taylor}, {Gibson}, {Peracaula}, {Martin},
  {Landecker}, {Brunt}, {Dewdney}, {Dougherty}, {Gray}, {Higgs}, {Kerton},
  {Knee}, {Kothes}, {Purton}, {Uyaniker}, {Wallace}, {Willis}, \&
  {Durand}}]{Taylor2003}
{Taylor}, A.~R., {Gibson}, S.~J., {Peracaula}, M., {et~al.} 2003, \aj, 125,
  3145, \dodoi{10.1086/375301}

\bibitem[{{Tielens} \& {Hollenbach}(1985)}]{Tielens1985}
{Tielens}, A.~G.~G.~M., \& {Hollenbach}, D. 1985, \apj, 291, 722,
  \dodoi{10.1086/163111}

\bibitem[{{Tiwari} {et~al.}(2021){Tiwari}, {Karim}, {Pound}, {Wolfire},
  {Jacob}, {Buchbender}, {G{\"u}sten}, {Guevara}, {Higgins}, {Kabanovic},
  {Pabst}, {Ricken}, {Schneider}, {Simon}, {Stutzki}, \&
  {Tielens}}]{Tiwari2021}
{Tiwari}, M., {Karim}, R., {Pound}, M.~W., {et~al.} 2021, \apj, 914, 117,
  \dodoi{10.3847/1538-4357/abf6ce}

\bibitem[{{Van De Putte} {et~al.}(2020){Van De Putte}, {Gordon}, {Roman-Duval},
  {Williams}, {Baes}, {Tchernyshyov}, {Lawton}, \& {Arab}}]{vanDePutte2020}
{Van De Putte}, D., {Gordon}, K.~D., {Roman-Duval}, J., {et~al.} 2020, \apj,
  888, 22, \dodoi{10.3847/1538-4357/ab557f}

\bibitem[{{van Leeuwen}(2007)}]{vanLeeuwen2007}
{van Leeuwen}, F. 2007, \aap, 474, 653, \dodoi{10.1051/0004-6361:20078357}

\bibitem[{{Wang} {et~al.}(2020){Wang}, {Bihr}, {Beuther}, {Rugel}, {Soler},
  {Ott}, {Kainulainen}, {Schneider}, {Klessen}, {Glover}, {McClure-Griffiths},
  {Goldsmith}, {Johnston}, {Menten}, {Ragan}, {Anderson}, {Urquhart}, {Linz},
  {Roy}, {Smith}, {Bigiel}, {Henning}, \& {Longmore}}]{Wang2020}
{Wang}, Y., {Bihr}, S., {Beuther}, H., {et~al.} 2020, \aap, 634, A139,
  \dodoi{10.1051/0004-6361/201935866}

\bibitem[{{Wilson} {et~al.}(2009){Wilson}, {Rohlfs}, \&
  {H{\"u}ttemeister}}]{Wilson2009}
{Wilson}, T.~L., {Rohlfs}, K., \& {H{\"u}ttemeister}, S. 2009, {Tools of Radio
  Astronomy}, \dodoi{10.1007/978-3-540-85122-6}

\bibitem[{{Witt} {et~al.}(1989){Witt}, {Stecher}, {Boroson}, \&
  {Bohlin}}]{witt1989}
{Witt}, A.~N., {Stecher}, T.~P., {Boroson}, T.~A., \& {Bohlin}, R.~C. 1989,
  \apjl, 336, L21, \dodoi{10.1086/185352}

\bibitem[{{Wolfire} {et~al.}(2022){Wolfire}, {Vallini}, \&
  {Chevance}}]{Wolfire2022}
{Wolfire}, M.~G., {Vallini}, L., \& {Chevance}, M. 2022, \araa, 60, 247,
  \dodoi{10.1146/annurev-astro-052920-010254}

\bibitem[{{Young} {et~al.}(2012){Young}, {Becklin}, {Marcum}, {Roellig}, {De
  Buizer}, {Herter}, {G{\"u}sten}, {Dunham}, {Temi}, {Andersson}, {Backman},
  {Burgdorf}, {Caroff}, {Casey}, {Davidson}, {Erickson}, {Gehrz}, {Harper},
  {Harvey}, {Helton}, {Horner}, {Howard}, {Klein}, {Krabbe}, {McLean}, {Meyer},
  {Miles}, {Morris}, {Reach}, {Rho}, {Richter}, {Roeser}, {Sandell}, {Sankrit},
  {Savage}, {Smith}, {Shuping}, {Vacca}, {Vaillancourt}, {Wolf}, \&
  {Zinnecker}}]{Young2012}
{Young}, E.~T., {Becklin}, E.~E., {Marcum}, P.~M., {et~al.} 2012, \apjl, 749,
  L17, \dodoi{10.1088/2041-8205/749/2/L17}

\end{thebibliography}
\bibliographystyle{aasjournal}




\end{document}